\documentclass{intech-hep}
\pdfoutput=1

\setcitestyle{numbers}

\usepackage{upgreek}
\usepackage{amsfonts}
\usepackage{amsbsy}
\usepackage{amscd}
\usepackage{bbm}
\usepackage{amssymb}

\newcommand{\bal}{\begin{align}}
\newcommand{\eal}{\end{align}}
\newcommand{\beqs}{\begin{subequations}}
\newcommand{\eeqs}{\end{subequations}}
\newcommand{\ec}{\end{center}}
\newcommand{\bec}{\begin{center}}

\newcommand{\eem}{\end{matrix}}
\newcommand{\bem}{\begin{matrix}}
\newcommand{\eeq}{\end{equation}}
\newcommand{\beq}{\begin{equation}}
\newcommand{\ba}{\begin{array}}
\newcommand{\ea}{\end{array}}
\newcommand{\bea}{\begin{eqnarray}}
\newcommand{\eea}{\end{eqnarray}}
\newcommand{\baq}{\begin{eqnarray}}
\newcommand{\eaq}{\end{eqnarray}}

\newcommand\eqs[2]{Eqs.~(\ref{#1}) and (\ref{#2})}

\newcommand\eqss[3]{Eqs.~(\ref{#1}), (\ref{#2}) and (\ref{#3})}

\newcommand{\ftn}{\footnotesize}

\newcommand{\ssz}{\scriptsize}
\newcommand{\TeV}{{\mbox{\rm TeV}}}

\newcommand{\GeV}{{\mbox{\rm GeV}}}
\newcommand{\eV}{{\mbox{\rm eV}}}

\newcommand{\sFref}[2]{Fig.~\ref{#1}-{\small\sf ({#2})}}
\newcommand{\sEref}[2]{Eq.~(\ref{#1}{\small\sf {#2}})}

\newcommand{\Eref}[1]{Eq.~(\ref{#1})}
\newcommand{\Sref}[1]{Sec.~\ref{#1}}
\newcommand{\Fref}[1]{Fig.~\ref{#1}}
\newcommand{\Tref}[1]{Table~\ref{#1}}
\newcommand{\cref}[1]{Ref.~\cite{#1}}

\newcommand{\etal}{{\it et al.\/}}

\def\dg{\dagger}
\def\llgm{\left\lgroup}
\def\rrgm{\right\rgroup}
\def\lf{\left(}
\def\rg{\right)}
\newcommand\vev[1]{\langle {#1} \rangle}
\newcommand{\Gr}{\ensuremath{\widetilde{G}}}

\newcommand{\Yg}{\ensuremath{Y_{\Gr}}}

\newcommand{\Vhi}{\ensuremath{\widehat V_{\rm HI}}}
\newcommand{\Hhi}{\ensuremath{\widehat H_{\rm HI}}}

\newcommand{\Khi}{\ensuremath{K}}

\newcommand{\Vhio}{\ensuremath{\widehat V_{\rm HI0}}}

\newcommand{\mP}{\ensuremath{m_{\rm P}}}
\newcommand{\Mpq}{\ensuremath{M_{\rm PS}}}
\newcommand{\mpq}{\ensuremath{m_{\rm PS}}}
\newcommand{\Mgut}{\ensuremath{M_{\rm GUT}}}
\newcommand{\Ggut}{\ensuremath{G_{\rm PS}}}
\newcommand{\Ms}{\ensuremath{M_{\rm S}}}

\def\openep{\leavevmode\hbox{\normalsize{\boldmath $\varepsilon$}}}
\def\openone{\leavevmode\hbox{\small1\kern-3.8pt\normalsize1}}

\newcommand{\ca}{\ensuremath{{\rm a}}}
\newcommand{\ck}{\ensuremath{c_\mathcal{R}}}
\newcommand{\kx}{\ensuremath{k_S}}
\newcommand{\kn}{\ensuremath{k_{H}}}

\newcommand{\Gsm}{\ensuremath{G_{\rm SM}}}

\newcommand{\Gsn}{\ensuremath{\Gamma_{\rm I}}}
\newcommand{\GNsn[1]}{\ensuremath{\Gamma_{{\rm I}{#1}\what\nu^c}}}
\newcommand{\Ghsn}{\ensuremath{\Gamma_{{\rm I}y}}}

\newcommand{\msn}{\ensuremath{m_{\rm I}}}

\newcommand{\hd}{{\ensuremath{H_d}}}
\newcommand{\hu}{{\ensuremath{H_u}}}

\newcommand{\ns}{\ensuremath{n_{\rm s}}}
\newcommand{\as}{\ensuremath{\alpha_{\rm s}}}

\newcommand{\Ve}{\ensuremath{\widehat{V}}}
\newcommand{\He}{\ensuremath{\widehat{H}}}
\newcommand{\Ne}{\ensuremath{\widehat{N}}}
\newcommand{\sni}{\ensuremath{\nu^c_i}}

\newcommand{\rhni}{\ensuremath{\nu^c_i}}
\newcommand{\wrhn[1]}{\ensuremath{\what\nu^c_{#1}}}
\newcommand{\mrh[1]}{\ensuremath{M_{{#1} \what\nu^c}}}
\newcommand{\mD[1]}{\ensuremath{m_{#1\rm D}}}
\newcommand{\mn[1]}{\ensuremath{m_{#1\nu}}}

\newcommand{\snH}{\ensuremath{\nu^c_H}}
\newcommand{\snHb}{\ensuremath{\bar\nu^c_H}}
\newcommand{\uH}{\ensuremath{u^c_H}}
\newcommand{\uHb}{\ensuremath{\bar u^c_H}}
\newcommand{\dH}{\ensuremath{d^c_H}}
\newcommand{\dHb}{\ensuremath{\bar d^c_H}}
\newcommand{\eH}{\ensuremath{e^c_H}}
\newcommand{\eHb}{\ensuremath{\bar e^c_H}}
\newcommand{\lH}{\ensuremath{\lambda_H}}
\newcommand{\lHb}{\ensuremath{\lambda_{\bar H}}}
\newcommand{\Hcc}{\ensuremath{H^c}}
\newcommand{\bHc}{\ensuremath{\bar H^c}}
\newcommand{\what}{\ensuremath{\widehat}}

\def\ve{\varepsilon}

\def\bbet{{\bar\beta}}
\def\al{{\alpha}}
\def\bt{{\beta}}

\def\n{\bar{n}}

\def\th{{\theta}}
\def\thb{{\bar\theta}}
\def\thn{{\theta_{\nu}}}

\newcommand{\Trh}{\ensuremath{T_{\rm rh}}}
\newcommand{\sg}{\ensuremath{h}}

\newcommand{\sgf}{\ensuremath{h_{\rm f}}}
\newcommand{\xsg}{\ensuremath{x_{h}}}

\newcommand{\ld}{\ensuremath{\lambda}}

\newcommand{\Ld}{\ensuremath{\Lambda}}

\newcommand{\se}{\ensuremath{\widehat h}}
\newcommand{\sex}{\ensuremath{\widehat{h}_*}}

\newcommand{\Gm[1]}{\ensuremath{\Gamma_{#1}}}

\newcommand{\eL}{{\ensuremath{\varepsilon_{L}}}}

\def\Ka{K\"{a}hler potential}

\def\FHI{nMHI~}

\newcommand{\Tr}{\mbox{\sf Tr}}
\newcommand{\diag}{\ensuremath{{\sf diag}}}
\newcommand{\im}{\ensuremath{{\sf Im}}}

\newcommand{\tr}{{\mbox{\sf\ssz T}}}

\newcommand\mtt[4]{\mbox{
$\llgm\bem #1 &#2 \cr #3& #4\eem\rrgm$}}

\newcommand{\bdhh}{{\ensuremath{\normalsize I{\kern-2.9pt H}}}}

\newcommand\lin[2]{\mbox{$\llgm\bem #1& #2\eem\rrgm$}}

\newcommand\mtn[9]{\ensuremath{\llgm\bem #1&#2&#3\cr #4&#5&#6 \cr #7&#8&#9\eem\rrgm}}

\booktitle{Open Questions in Cosmology}%

\chaptertitle{Leptogenesis and Neutrino Masses in an Inflationary SUSY Pati-Salam  Model}

\authors{C. Pallis and N. Toumbas \\ {\normalsize\sl Department of Physics, University of Cyprus, \\
P.O. Box 20537, CY-1678 Nicosia, CYPRUS}}

\webadress{http://dx.doi.org/10.5772/b.bookid.chapterid}

\begin{document}

\maketitle

\section*{Abstract}

We implement the mechanism of non-thermal leptogenesis in the
framework of an inflationary  model based on a
\emph{supersymmetric} (SUSY) Pati-Salam \emph{Grand Unified
Theory} (GUT). In particular, we show that inflation is driven by
a quartic potential associated with the Higgs fields involved in
the spontaneous GUT symmetry breaking, in the presence of a
non-minimal coupling of the inflaton field to gravity. The
inflationary model relies on renormalizable superpotential terms
and does not lead to overproduction of magnetic monopoles. It is
largely independent of one-loop radiative corrections, and it can
be consistent with current observational data on the inflationary
observables, with the GUT symmetry breaking scale assuming its
SUSY value. Non-thermal leptogenesis is realized by the
out-of-equilibrium decay of the two lightest \emph{right-handed}
(RH) neutrinos, which are produced by the inflaton decay.
Confronting our scenario with the current observational data on
light neutrinos, the GUT prediction for the heaviest Dirac
neutrino mass, the baryon asymmetry of the universe and the
gravitino limit on the reheating temperature, we constrain the
masses of the RH neutrinos in the range $(10^{10}-10^{15})~\GeV$
and the Dirac neutrino masses of the two first generations to
values between $0.1$ and $20~\GeV$.

{\bf Keywords:} {\sf  Baryogenesis, Inflation, Grand Unified
Theories}

{\bf PACs Codes:} {\sf 98.80.Cq, 11.30.Qc, 11.30.Er, 11.30.Pb,
12.60.Jv}

\section{Introduction}

One of the most promising and well-motivated mechanisms for the
generation of the \emph{Baryon Asymmetry of the Universe} (BAU) is
via an initial generation of a lepton asymmetry, which can be
subsequently converted to BAU through sphaleron effects -- see
e.g. \cref{baryo1, baryo}. \emph{Non-Thermal Leptogenesis} (nTL)
\cite{inlept,inlept1} is a variant of this proposal, in which the
necessitated departure from equilibrium is achieved by
construction. Namely, the \emph{right-handed} (RH) neutrinos,
$\rhni$, whose decay produces the lepton asymmetry, are
out-of-equilibrium at the onset, since their masses are larger
than the reheating temperature. Such a set-up can be achieved by
the direct production of $\rhni$ through the inflaton decay, which
can also take place out-of-equilibrium. Therefore, such a
leptogenesis paradigm largely depends on the inflationary stage,
which it follows.

In a recent paper \cite{nmH} -- for similar attempts, see
\cref{sm1,nmSUGRA2,jones2} --, we investigate an inflationary
model where a \emph{Standard Model} (SM) singlet component of the
Higgs fields involved in the spontaneous breaking of a
\emph{supersymmetric} (SUSY) \emph{Pati-Salam} (PS) \emph{Grand
Unified Theory} (GUT) can produce inflation of chaotic-type, named
\emph{non-minimal Higgs Inflation} (nMHI), since there is a
relatively strong non-minimal coupling of the inflaton field to
gravity \cite{ajones,lee,linde1,linde2}. This GUT provides a
natural framework to implement our leptogenesis scenario, since
the presence of the $SU(2)_{\rm R}$ gauge symmetry predicts the
existence of three $\rhni$. In its simplest realization this GUT
leads to third family \emph{Yukawa unification} (YU), and does not
suffer from the doublet-triplet splitting problem since both Higgs
doublets are contained in a bidoublet other than the GUT scale
Higgs fields. Although this GUT is not completely unified -- as,
e.g., a GUT based on the $SO(10)$ gauge symmetry group -- it
emerges in standard weakly coupled heterotic string models
\cite{leontaris} and in recent D-brane constructions
\cite{branes}.

The inflationary model relies on renormalizable superpotential
terms and does not lead to overproduction of magnetic monopoles.
It is largely independent of the one-loop radiative corrections
\cite{cw}, and it can become consistent with the fitting
\cite{wmap} of the seven-year data of the \emph{Wilkinson
Microwave Anisotropy Probe Satellite} (WMAP7) combined with the
\emph{baryon-acoustic oscillation} (BAO) and the measurement of
the \emph{Hubble constant} ($H_0$). At the same time the GUT
symmetry breaking scale attains its SUSY value and the $\mu$
problem of the \emph{Minimal SUSY SM} (MSSM) is resolved via a
Peccei-Quinn (PQ) symmetry, solving also the strong CP problem.
Inflation can be followed by non-thermal leptogenesis, compatible
with the gravitino ($\Gr$) limit \cite{gravitino, brand, kohri} on
the reheating temperature, leading to efficient baryogenesis. In
\cref{nmH} we connect non-thermal leptogenesis with neutrino data,
implementing a two-generation formulation of the see-saw
\cite{seesaw,seesaw1,seesaw2} mechanism and imposing extra
restrictions from the data on the light neutrino masses and the
GUT symmetry on the heaviest Dirac neutrino mass. There we
\cite{nmH} assume that the mixing angle between the first and
third generation, $\theta_{13}$, vanishes. However, the most
updated \cite{valle12, lisi12} analyses of the low energy neutrino
data suggest that non-zero values for $\theta_{13}$ are now
preferred, while the zero value can be excluded at $8$ standard
deviations. Therefore, a revision of our results, presented in
\cref{nmH}, is worth pursuing.

The three-generation implementation of the see-saw mechanism is
here adopted, following a bottom-up approach, along the lines of
\cref{branco, frigerio, senoguz, nMCI}. In particular, we use as
input parameters the low energy neutrino observables considering
several schemes of neutrino masses. Using also the third
generation Dirac neutrino mass predicted by the PS GUT, assuming a
mild hierarchy for the two residual generations and imposing the
restriction from BAU, we constrain the masses of $\rhni$'s and
the residual neutrino Dirac mass spectrum. Renormalization group
effects \cite{nMCI, running} are also incorporated in our
analysis.

We present the basic ingredients of our model in Sec.~\ref{fhim}.
In Sec.~\ref{fhi} we describe the inflationary potential and
derive the inflationary observables. In Sec.~\ref{pfhi} we outline
the mechanism of non-thermal leptogenesis, while in \Sref{cont} we
exhibit the relevant imposed constraints and restrict the
parameters of our model. Our conclusions are summarized in
Sec.~\ref{con}. Throughout the text, we use natural units for
Planck's and Boltzmann's constants and the speed of light
($\hbar=c=k_{\rm B}=1$); the subscript of type $,\chi$ denotes
derivation \emph{with respect to} (w.r.t) the field $\chi$ (e.g.,
$_{,\chi\chi}=\partial^2/\partial\chi^2$); charge conjugation is
denoted by a star and $\log~[\ln]$ stands for logarithm with basis
$10~[e]$.

\section{The Pati-Salam SUSY GUT Model}\label{fhim}

In this section, we present the particle content (\Sref{fhim1}), the
structure of the superpotential and the \Ka (\Sref{fhim2}) and describe the
SUSY limit (\Sref{fhim3}) of our model.

\subsection{Particle Content}\label{fhim1}

We focus on a SUSY PS GUT model described in detail in
Ref.~\cite{jean, nmH}. The representations and the transformation
properties of the various superfields under
$G_{\rm PS}=SU(4)_{\rm C}\times SU(2)_{\rm L}\times SU(2)_{\rm
R}$, their decomposition under ${G_{\rm SM}}= SU(3)_{\rm C}\times
SU(2)_{\rm L}\times U(1)_{Y}$, as well as their extra global
charges are presented in Table~\ref{tab1}.

The $i$th generation $(i=1,2,3)$ \emph{left-handed} (LH) quark and
lepton superfields, $u_{i\ca}$, $d_{i\ca}$ ($\ca=1,2,3$ is a color
index), $e_i$ and $\nu_i$ are accommodated in a superfield $F_i$.
The LH antiquark and antilepton superfields $u^c_{i\ca}$,
$d_{i\ca}^c$, $e^c_i$ and $\sni$ are arranged in another
superfield $F^c_i$. The gauge symmetry $G_{\rm PS}$ can be
spontaneously broken down to $G_{\rm SM}$ through v.e.vs which the
superfields $H^c$ and $\bar H^c$ acquire in the directions
$\nu^c_H$ and $\bar\nu^c_H$. The model also contains a gauge
singlet $S$, which triggers the breaking of $G_{\rm PS}$, as well
as an $SU(4)_{\rm C}$ {\bf 6}-plet $G$, which splits into
$g_\ca^c$ and $\bar{g}^c_\ca$ under $G_{SM}$ and gives
\cite{leontaris} superheavy masses to $d^c_{H\ca}$ and
$\bar{d}^c_{H\ca}$. In the simplest realization of this model
\cite{leontaris, jean}, the electroweak doublets $\hu$ and $\hd$,
which couple to the up and down quarks respectively, are
exclusively contained in the bidoublet superfield $\bdhh$.

\renewcommand{\arraystretch}{1.2}

\begin{table}[!t]
\begin{center}
\begin{tabular}{|ccccccc|}
\hline {\sc Super-}&{\sc Represe-}&{\sc Trasfor-}&{\sc
Decompo-}&\multicolumn{3}{c|}{ {\sc Global} }
\\
\multicolumn{1}{|c}{\sc fields}\hspace*{0.2cm}&{\sc ntations}
&{\sc mations} &{\sc sitions} &\multicolumn{3}{c|}{ {\sc Charges}}
\\
\multicolumn{1}{|c}{}&{\sc under $G_{\rm PS}$}\hspace*{0.2cm}
&{\sc under $G_{\rm PS}$}\hspace*{0.2cm}&{\sc under $G_{\rm
SM}$}\hspace*{0.2cm} &{$~~~~R~~~$} &{PQ~~~~} &{$\mathbb{Z}^{\rm
mp}_2$}
\\\hline \hline
\multicolumn{7}{|c|}{\sc Matter Superfields}
\\ \hline
{$F_i$} &{$({\bf 4, 2, 1})$}&$F_iU_{\rm L}^{\dagger}U^\tr_{\rm C}$
& $Q_{i\rm a}({\bf 3,2},1/6)$&$1$ &
$-1$ &$-$\\
&&& $L_i({\bf 1,2},-1/2)$&&&\\
{$F^c_i$} & {$({\bf \bar 4, 1, 2})$}&$U_{\rm C}^\ast U_{\rm
R}^\ast
F^c_i$ &$u_{i\rm a}^c({\bf\bar 3,1},-2/3)$&{ $1$ }&{$0$}&{$-$}\\
&&& $d_{i\rm a}^c({\bf\bar 3,1},1/3)$&&&\\
&&& $\nu_i^c({\bf 1,1},0)$&&&\\
&&& $e_i^c({\bf 1,1},1)$&&&\\\hline%
\multicolumn{7}{|c|}{\sc Higgs Superfields}\\ \hline {$H^c$}
&{$({\bf \bar 4, 1, 2})$}& $U_{\rm C}^\ast U_{\rm R}^\ast H^c$
&$u_{H{\rm a}}^c({\bf \bar 3,1},-2/3)$&{$0$}&{$0$} & {$+$}\\
&&& $d_{H{\rm a}}^c({\bf \bar 3,1},1/3)$&&&\\
&&& $\nu_H^c({\bf 1,1},0)$&&&\\
&&& $e_H^c({\bf 1,1},1)$&&&\\
{$\bar H^c$}&$({\bf 4, 1, 2})$& $\bar{H}^cU^\tr_{\rm R}
U^\tr_{\rm C}$&$\bar u_{H{\rm a}}^c({\bf 3,1},2/3)$&{$0$}&{$0$}&{$+$} \\
&&& $\bar d_{H\rm a}^c({\bf 3,1},-1/3)$&&&\\
&&& $\bar \nu_H^c({\bf 1,1},0)$&&&\\
&&& $\bar e_H^c({\bf 1,1},-1)$&&&\\
{$S$} & {$({\bf 1, 1, 1})$}&$S$ &$S ({\bf 1,1},0)$&$2$ &$0$ &$+$ \\
{$G$} & {$({\bf 6, 1, 1})$}&$U_{\rm C}GU^\tr_{\rm C}$ &$\bar
g_\ca^c({\bf 3,1},-1/3)$&$2$ &$0$ &$+$\\ &&& $g_\ca^c({\bf \bar
3,1},1/3)$&&&\\\hline
{$\bdhh$} & {$({\bf 1, 2, 2})$}&$U_{\rm L}\bdhh U^\tr_{\rm R}$ &$H_u({\bf 1,2},1/2)$&$0$ &$1$ &$+$\\
&&& $H_d({\bf 1,2}, -1/2)$&&&\\ \hline
{$P$} &{$({\bf 1, 1, 1})$}& $P$ &$P ({\bf 1,1},0)$&{$1$}&{$-1$} & {$+$} \\
{$\bar P$}&$({\bf 1, 1, 1})$& $\bar P$&$\bar P ({\bf
1,1},0)$&{$0$}&{$1$}&{$+$}\\ \hline
\end{tabular}
\end{center}
\caption{\sl\small The representations, the transformations under
$G_{\rm PS}$, the decompositions under $G_{\rm SM}$ as well as the
extra global charges of the superfields of our model. Here $U_{\rm
C}\in SU(4)_{\rm C}$, $U_{\rm L}\in SU(2)_{\rm L}$, $U_{\rm R}\in
SU(2)_{\rm R}$ and $\tr,~\dagger$ and $\ast$ stand for the
transpose, the hermitian conjugate and the complex conjugate of a
matrix respectively.}\label{tab1}
\end{table}

In addition to $G_{\rm PS}$, the model possesses two global $U(1)$
symmetries, namely a PQ and an R symmetry, as well as a discrete
$\mathbb{Z}_2^{\rm mp}$ symmetry (`matter parity') under which
$F$, $F^c$ change sign. The last symmetry forbids undesirable
mixings of $F$ and $\bdhh$ and/or $F^c$ and $H^c$ and ensures the
stability of the \emph{lightest SUSY particle} (LSP). The imposed
$U(1)$ R symmetry, $U(1)_R$, guarantees the linearity of the
superpotential w.r.t the singlet $S$. Finally the $U(1)$ PQ
symmetry, $U(1)_{\rm PQ}$, assists us to generate the $\mu$-term
of the MSSM. The PQ breaking occurs at an intermediate scale
through the v.e.vs of $P$, $\bar{P}$, and the $\mu$-term is
generated via a non-renormalizable coupling of $P$ and $\bdhh$.
Following \cref{jean}, we introduce into the scheme quartic
(non-renormalizable) superpotential couplings of $\bar{H}^c$ to
$F^c_i$, which generate intermediate-scale masses for the $\sni$
and, thus, masses for the light neutrinos, $\nu_i$, via the seesaw
mechanism \cite{seesaw,seesaw1,seesaw2}. Moreover, these couplings
allow for the decay of the inflaton into  $\nu^c_i$, leading to a
reheating temperature consistent with the $\Gr$ constraint with
more or less natural values of the parameters. As shown finally in
\cref{jean}, the proton turns out to be practically stable in this
model.

\subsection{Superpotential and K\"ahler Potential}\label{fhim2}

The superpotential $W$ of our model splits into three parts:
\beq W=W_{\rm MSSM}+W_{\rm PQ}+W_{\rm HPS}, \label{Wtotal}\eeq
which are analyzed in the following.

\begin{itemize}

\item $W_{\rm MSSM}$ is the part of $W$ which contains the usual
terms -- except for the $\mu$ term -- of the MSSM, supplemented by
Yukawa interactions among the left-handed leptons and $\sni$:
\bea \nonumber  W_{\rm MSSM} = y_{ij} F_i \bdhh F^c_j= \hspace*{5.cm} \\
 = y_{ij}\lf \hd^\tr\openep {L}_ie^c_j -\hu^\tr \openep L_i\nu^c_j
+\hd^\tr \openep Q_{i\ca}d^c_{j\ca} -\hu^\tr \openep {Q}_{i\ca}
u^c_{j\ca} \rg,\>\>\>\mbox{with}\>\>\>\openep=\llgm\bem0&1\cr-1
&0\eem\rrgm\cdot \label{wmssm}\eea
Here $Q_{i\ca}=\lin{u_{i\ca}}{d_{i\ca}}^\tr$ and
$L_{i}=\lin{\nu_i}{e_i}^\tr$ are the $i$-th generation $SU(2)_{\rm
L}$ doublet LH quark and lepton superfields respectively.
Summation over repeated color and generation indices is assumed.
Obviously the model predicts YU at $\Mgut$ since the fermion
masses per family originate from a unique term of the PS GUT. It
is shown \cite{oliveira, shafi} that exact third family YU
combined with non-universalities in the gaugino sector and/or the
scalar sector can become consistent with a number of
phenomenological and cosmological low-energy requirements. On the
other hand, it is expected on generic grounds that the predictions
of this simple model for the fermion masses of the two lighter
generations are not valid. Usually this difficulty can be avoided
by introducing \cite{oliveira1} an abelian symmetry which
establishes a hierarchy between the flavor dependent couplings.
Alternatively, the present model can be augmented \cite{lpnova}
with other Higgs fields so that $\hu$ and $\hd$ are not
exclusively contained in $\bdhh$, but receive subdominant
contributions from other representations too. As a consequence, a
moderate violation of exact YU can be achieved, allowing for an
acceptable low-energy phenomenology, even with universal boundary
conditions for the soft SUSY breaking terms. However, we prefer
here to work with the simplest version of the PS model, using the
prediction of the third family YU in order to determine the
corresponding Dirac neutrino mass -- see \Sref{cont1}.

\item $W_{\rm PQ}$, is the part of $W$ which is relevant for the
spontaneous breaking of $U(1)_{\rm PQ}$ and the generation of the
$\mu$ term of the MSSM. It is given by
\beq\label{Wpq} W_{\rm PQ}=\ld_{\rm PQ} \frac{P^2
\bar{P}^2}{\Ms}-\ld_\mu \frac{P^2}{2\Ms}\Tr\left(\bdhh\openep
\bdhh^{\tr}\openep\right),  \eeq
where $\Ms\simeq 5\cdot 10^{17}~{\rm GeV}$ is the String scale.
The scalar potential, which is generated by the first term in the
RHS of \Eref{Wpq}, after gravity-mediated SUSY breaking, is studied
in \cref{rsym,jean}. For a suitable choice of parameters, the
minimum lies at $\vert \vev{P}\vert = \vert \vev{\bar{P}}\vert
\sim \sqrt{m_{3/2}M_{\rm S}}$. Hence, the PQ symmetry breaking
scale is of order $\sqrt{m_{3/2} \Ms} \simeq \lf10^{10} -
10^{11}\rg~\GeV$. The $\mu$-term of the MSSM is generated from
the second term of the RHS of \Eref{Wpq} as follows:
\beq -\ld_\mu \frac{\vev{P}^2}{2\Ms}\Tr\left(\bdhh\openep
\bdhh^{\tr}\openep\right)=\mu\hd^\tr\openep\hu\>\>\Rightarrow
\>\>\mu\simeq\ld_\mu {\vev{P}^2\over\Ms},\eeq
which is of the right magnitude if $\lambda_\mu\sim(0.001-0.01)$.
Let us note that $V_{\rm PQ}$ has an additional local minimum at
$P=\bar{P}=0$, which is separated from the global PQ minimum by a
sizable potential barrier, thus preventing transitions from the
trivial to the PQ vacuum. Since this situation persists at all
cosmic temperatures after reheating, we are obliged to assume
that, after the termination of nMHI, the system emerges with the
appropriate combination of initial conditions so that it is led
\cite{curvaton} in the PQ vacuum.

\item $W_{\rm HPS}$, is the part of $W$ which is relevant for
nMHI, the spontaneous breaking of $G_{\rm PS}$ and the generation
of intermediate Majorana [superheavy] masses for $\sni$ [$\dH$ and
$\dHb$]. It takes the form
\beq W_{\rm HPS}=\ld S \lf\bHc\Hcc -\Mpq^2\rg +\lH H^{c\tr} G
\openep H^c + \lHb \bar{H}^{c} \bar G \openep \bar{H}^{c\tr}+
\ld_{i\nu^c} \frac{\lf\bar{H}^c F_i^c\rg^2}{\Ms},\label{Whi} \eeq
where $\Mpq$ is a superheavy mass scale related to $\Mgut$ -- see
\Sref{fhi2} -- and $\bar G$ is the dual tensor of $G$. The
parameters $\ld$ and $\Mpq$ can be made positive by field
redefinitions.

\end{itemize}

According to the general recipe \cite{linde1, linde2}, the
implementation of \FHI within SUGRA requires the adoption of a
K\"ahler potential, $\Khi$, of the following type
\beq  \Khi=-3\mP^2\ln \lf 1- {H^{c\dg}\Hcc\over3\mP^2}-{\bHc\bar
H^{c\dg}\over3\mP^2}-{\Tr \lf G^\dg G\rg \over6\mP^2}-
{|S|^2\over3\mP^2}+{\kx}{|S|^4\over3\mP^4}+{\kn\over2\mP^2}\lf
\bHc\Hcc +{\rm h.c.}\rg\rg,\label{minK}\eeq
where $\mP=2.44\cdot10^{18}~\GeV$ is the reduced Planck scale and
the complex scalar components of the superfields $\Hcc, \bHc, G$
and $S$ are denoted by the same symbol. The coefficients $\kx$ and
$\kn$ are taken real. From \Eref{minK} we can infer that we adopt
the standard quadratic non-minimal coupling for Higgs-inflaton,
which respects the gauge and global symmetries of the model. This
non-minimal coupling of the Higgs fields to gravity is transparent
in the Jordan frame. We also added the fifth term in the RHS of
\Eref{minK} in order to cure the tachyonic mass problem
encountered in similar models \cite{lee, linde1, linde2} -- see
\Sref{fhi1}. In terms of the components of the various fields,
$\Khi$ in \Eref{minK} reads
\beqs\beq \Khi=-3\mP^2\ln \lf1-
{\phi^\al\phi^{*\bar\al}\over3\mP^2}+{\kx}{|S|^4\over3\mP^4}+
{\kn\over2\mP^2}\lf\snH\snHb+\eH\eHb+\uH\uHb+\dH\dHb+{\rm
h.c.}\rg\right) \label{Kcom} \eeq
with \beq \phi^\al=\snH, \snHb, \eH, \eHb, \uH, \uHb, \dH, \dHb,
g^c, \bar g^c ~~\mbox{and}~~ S \label{fas}\eeq\eeqs and summation
over the repeated Greek indices is
implied.

\subsection{The SUSY Limit}\label{fhim3}

In the limit where $\mP$ tends to infinity, we can obtain the SUSY
limit of the SUGRA potential. Assuming that the SM non-singlet components
vanish, the F-term potential in this limit, $V_{\rm F}$, turns out to be
\beqs \beq\label{VF}V_{\rm F} = \ld^2 \left\vert\snHb \snH -
\Mpq^2\right\vert^2 + \ld^2 \vert S \vert^2 \left(|\snH|^2
+|\snHb|^2\right),\eeq
while the D-term potential is
\beq V_{\rm D} =  {5g^2\over 16} \lf|\snH|^2 -
|\snHb|^2\rg^2 .
\label{Vd1}\eeq\eeqs
Restricting ourselves to the D-flat direction $|\snH|= |\snHb|$,
we find from $V_{\rm F}$ that the SUSY vacuum
lies at
\beq
\vev{S}\simeq0\>\>\>\mbox{and}\>\>\>\left|\vev{\snH}\right|=\left|\vev{\snHb}\right|=\Mpq.
\label{vevs} \eeq
Therefore, $W_{\rm HPS}$ leads to spontaneous breaking of
$\Ggut$. As we shall see in \Sref{fhi}, the same superpotential, $W_{\rm HPS}$, gives rise
to a stage of \FHI. Indeed, along the
D-flat direction $|\snH|=|\snHb|\gg \Mpq$ and $S=0$, $V_{\rm
SUSY}$ tends to a quartic potential, which can be
employed in conjunction with $\Khi$ in \Eref{minK} for the
realization of \FHI along the lines of \cref{linde2}.

It should be mentioned that soft SUSY breaking and instanton
effects explicitly break $U(1)_R\times U(1)_{\rm PQ}$ to
$\mathbb{Z}_2\times \mathbb{Z}_6$. The latter symmetry is
spontaneously broken by $\vev{P}$ and $\vev{\bar{P}}$. This would
lead to a domain wall problem if the PQ transition took place
after nMHI. However, as we already mentioned above, $U(1)_{\rm
PQ}$ is assumed already broken before or during nMHI. The final
unbroken symmetry of the model is $\Gsm \times \mathbb{Z}_2^{\rm
mp}$.

\section{The Inflationary Scenario}\label{fhi}

Next we outline the salient features of our inflationary scenario
(\Sref{fhi1}) and calculate a number of observable quantities in
Sec.~\ref{fhi2}.

\subsection{Structure of the Inflationary
Potential}\label{fhi1}

At tree-level the \emph{Einstein Frame} (EF) SUGRA
potential, $\Vhi$, is given by
\cite{linde1}
\beqs\beq \Vhi=e^{\Khi/\mP^2}\left(K^{\al\bbet}{\rm F}_\al {\rm
F}_\bbet-3\frac{\vert W_{\rm HPS}\vert^2}{\mP^2}\right) +
{1\over2}g^2 \sum_a D_a D_a, \label{Vsugra} \eeq
where $g$ is the unified gauge coupling constant and the summation
is applied over the $21$ generators $T_a$ of the PS gauge group --
see \cref{nmH}. Also, we have
\beq K_{\al\bbet}={\Khi_{,\phi^\al\phi^{*\bbet}}},\>\>
K^{\bbet\al}K_{\al\bar \gamma}=\delta^\bbet_{\bar \gamma},\>\>{\rm
F}_\al=W_{{\rm HPS},\phi^\al} +K_{,\phi^\al}W_{\rm
HPS}/\mP^2\>\>\mbox{and}\>\>D_a=\phi_\al\lf T_a\rg^\al_\bbet
K^\bbet \eeq\eeqs
The $\phi^\al$'s are given in \Eref{fas}. If we parameterize
the SM singlet components of $\Hcc$ and $\bHc$ by
\beq\label{hpar}
\snH=he^{i\th}\cos\theta_\nu/\sqrt{2}\>\>\>\mbox{and}\>\>\>\snHb=he^{i\thb}\sin\theta_\nu/\sqrt{2},\eeq
we can easily deduce that a D-flat direction occurs at
\beq\label{inftr}
\th=\thb=0,\>\thn={\pi/4}\>\>\>\mbox{and}\>\>\>\eH=\eHb=\uH=\uHb=\dH=\dHb=g^c=\bar
g^c=0.\eeq
Along this direction, the D-terms in \Eref{Vsugra} -- and, also,
$V_{\rm D}$ in \Eref{Vd1} -- vanish, and so $\Vhi$ takes the form
\beq \label{Vhi}\Vhi
=\mP^4\frac{\ld^2(\xsg^2-4\mpq^2)^2}{16f^2}\eeq with \beq
\label{Vhi1} f=1+\ck
\xsg^2,\>\>\>\mpq={\Mpq\over\mP},\>\>\>\xsg={h\over\mP}~~
\mbox{and}\>\>\>\ck=-\frac{1}{6}+\frac{\kn}{4}\cdot\eeq
From \Eref{Vhi}, we can verify that for $\ck\gg1$ and $\mpq\ll1$,
$\Vhi$ takes a form suitable for the realization of nMHI, since it
develops a plateau -- see also \Sref{fhi2}. The (almost) constant
potential energy density $\Vhio$ and the corresponding Hubble
parameter $\He_{\rm HI}$ (along the trajectory in \Eref{inftr})
are given by
\beq \Vhio=
{\ld^2\sg^4\over16f^2}\simeq{\ld^2\mP^4\over16\ck^2}\>\>\>\mbox{and}\>\>\>
\He_{\rm
HI}={\Vhio^{1/2}\over\sqrt{3}\mP}\simeq{\ld\mP\over4\sqrt{3}\ck}\,\cdot\label{Vhio}\eeq

We next proceed to check the stability of the trajectory in
\Eref{inftr} w.r.t the fluctuations of the various fields. To this
end, we expand them in real and imaginary parts as follows
\bea X= {x_1+ix_2\over\sqrt{2}},\>\>\>\bar X= {\bar x_1+i\bar
x_2\over\sqrt{2}}\>\>\>\mbox{where}\>\>\>X=\eH,\uH,\dH, g^c
\>\>\>\mbox{and}\>\>\>x=e, u, d, g~. \label{cannor} \eea
Notice that the field $S$ can be rotated to the
real axis via a suitable R transformation. Along the trajectory in
\Eref{inftr} we find
\beq \lf K_{\al\bbet}\rg=\diag\lf {M_K\over
f^2},\underbrace{{1\over f},...,{1\over f}}_{3+6\cdot3 ~\mbox{\ftn
times}}\rg~~\mbox{with}~~
M_K=\mtt{\kappa}{\bar\kappa}{\bar\kappa}{\kappa},
\>\>\bar\kappa={3\ck^2\xsg^2}\>\>\>\mbox{and}\>\>\>
\kappa=f+\bar\kappa.\label{Sni1} \eeq
To canonically normalize the fields $\snH$ and $\snHb$, we first
diagonalize the matrix $M_K$. This can be achieved via a
similarity transformation involving an orthogonal matrix $U_K$ as
follows:
\beq \label{diagMk} U_K M_K U_K^\tr =\diag\lf\bar
f,f\rg,\>\>\>\mbox{where}\>\>\>\bar
f=f+6\ck^2\xsg^2\>\>\>\mbox{and}\>\>\>
U_K={1\over\sqrt{2}}\mtt{1}{1}{-1}{1}. \eeq
Utilizing $U_K$, the kinetic terms of the various fileds can be
brought into the following form
\beq K_{\al\bbet}\dot\phi^\al \dot\phi^{*\bbet}= {\bar f\over
2f^2}\lf\dot h ^2+{1\over2}h^2\dot\theta^2_+ \rg+{h^2\over
2f}\lf{1\over2}\dot\theta^2_- +\dot\theta^2_\nu
\rg+\frac{1}{2f}\dot\chi_\al\dot\chi_\al=\frac12\dot{\widehat
h}^2+\frac12\dot{\widehat \psi_\al}\dot{\widehat\psi_\al},
\label{Snik}\eeq
where $\th_{\pm}=\lf\bar\th\pm\th\rg/\sqrt{2}$,  $\chi_\al=x_1,
x_2, \bar x_1, \bar x_2, S$ and $\psi_\al=\th_+, \th_-,\th_\nu,
\chi_\al$ and the dot denotes derivation w.r.t the cosmic time,
$t$. In the last line, we introduce the EF canonically normalized
fields, $\what h$ and $\widehat \psi$, which can be obtained as
follows -- cf. \cref{linde1,linde2,nmN,nmH}:
\beq \label{VJe} \frac{d\se}{dh}=J={\sqrt{\bar f}\over
f},\>\>\widehat \theta_+ ={Jh\theta_+\over\sqrt{2}},\>\>\widehat
\theta_- ={h\theta_-\over\sqrt{2f}},\>\>\widehat \theta_\nu =
\frac{\sg}{\sqrt{f}}\lf\theta_\nu-{\pi\over4}\rg\>\>\mbox{and}\>\>\widehat
\chi_\al =\frac{\chi_\al}{\sqrt{f}} \cdot\eeq
%
Taking into account the approximate expressions for $\dot\sg$, $J$
and the slow-roll parameters $\widehat\epsilon, \,\widehat\eta$,
which are displayed in \Sref{fhi2}, we can verify that, during a
stage of slow-roll inflation,  $\dot{\widehat \th_+}\simeq J\sg\dot
\th_+/\sqrt{2}$  since $J\sg\simeq\sqrt{6}\mP$,  $\dot{\widehat
\th_-}\simeq \sg\dot \th_-/\sqrt{2f}$ and  $\dot{\widehat
\th_\nu}\simeq \sg\dot \th_\nu/\sqrt{f}$  since
$h/\sqrt{f}\simeq\mP/\sqrt{\ck}$. On the other hand, we can show
that $\dot{\widehat\chi_\al}\simeq\dot \chi_\al/\sqrt{f}$, since the quantity $\dot
f/2f^{3/2}\chi_\al$, involved in relating $\dot \chi_\al$ to $\dot{\widehat
\chi_\al}$, turns out to be negligibly small
compared with $\dot{\what\chi_\al}$. Indeed, the $\what
\chi_\al$'s acquire effective masses $m_{\what \chi_\al}\gg \Hhi$
-- see below -- and therefore enter a phase of oscillations about
$\what \chi_\al=0$ with decreasing amplitude. Neglecting the
oscillatory part of the relevant solutions, we find
\beq \chi\simeq\what
\chi_{\al0}\sqrt{f}e^{-2\Ne/3}\>\>\mbox{and}\>\>\>
\dot{\what\chi_\al}\simeq-2\chi_{\al0}\sqrt{f}\Hhi\what\eta_{\chi_\al}
e^{-2\Ne/3},\label{xdx}\eeq
where $\what \chi_{\al0}$ represents the initial amplitude of the
oscillations, $\what\eta_{\chi_\al}=m^2_{\what\chi_\al}/3\Hhi$ and
we assume $\dot{\what \chi_\al}(t=0)=0$. Taking into account the
approximate expressions for $\dot\sg$ and the slow-roll parameter
$\widehat\epsilon$ in \Sref{fhi2}, we find
\beq -{\dot f/2f^{3/2}}\chi_\al=\lf\ck\what\epsilon\He_{\rm HI}^2/
m_{\what \chi_\al}^2\rg \dot{\what\chi_\al}\ll
\dot{\what\chi_\al}.\label{ff32}\eeq

Having defined the canonically normalized scalar fields, we can
proceed in investigating the stability of the inflationary
trajectory of \Eref{inftr}. To this end, we expand $\Vhi$ in
\Eref{Vsugra} to quadratic order in the fluctuations around the
direction of \Eref{inftr}, as described in detail in \cref{nmH}.
In \Tref{tab2} we list the eigenvalues of the mass-squared
matrices
\beq
M^2_{\al\bt}=\left.{\partial^2\Vhi\over\partial\what\psi_\al\partial\what\psi_\beta}\right|_{\mbox{\Eref{inftr}}}
\mbox{with}~~\psi_\al=\th_+, \th_-,\th_\nu, x_{1}, x_{2},{\bar
x}_{1},{\bar x}_{2}~~\mbox{and}\>\>S\eeq
involved in the expansion of $\Vhi$. We arrange our findings into
three groups: the SM singlet sector, $S-\snH-\snHb$, the sector
with the $\uH,\uHb$ and the $\eH,\eHb$ fields which are related
with the broken generators of $\Ggut$ and the sector with the
$\dH, \dHb$ and the $g^c,\bar g^c$ fields.  Upon diagonalization
of the relevant matrices we obtain the following mass eigenstates:
\beq\widehat x_{1\pm}={1\over\sqrt{2}}\lf\widehat{\bar
x}_1\pm\widehat x_1\rg\>\>\>\mbox{and}\>\>\>\widehat
x_{2\mp}={1\over\sqrt{2}}\lf\widehat{\bar x}_2\mp\widehat
x_2\rg\>\>\>\mbox{with}\>\>\>x=u,e,d\>\>\mbox{and}\>\>g.\eeq

As we observe from the relevant eigenvalues, no instability -- as
the one found in \cref{nmN} -- arises in the spectrum. In
particular, it is evident that $\kx\gtrsim1$ assists us to achieve
$m_{\widehat S}^2>0$ -- in accordance with the results of
\cref{linde2}. Moreover, the D-term contributions to $m^2_{\what
\th_\nu}$ and $m^2_{\what u-}$ -- proportional to the gauge
coupling constant $g\simeq0.7$ -- ensure the positivity of these
masses squared. Finally the masses that the scalars $\what
d_{1,2}$ acquire from the second and third term of the RHS of
\Eref{Whi} lead to the positivity of $m_{\what d-}^2$ for $\lH$ of
order unity. We have also numerically verified that the masses of
the various scalars remain greater than the Hubble parameter
during the last $50-60$ e-foldings of nMHI, and so any
inflationary perturbations of the fields other than the inflaton
are safely eliminated.

\renewcommand{\arraystretch}{1.4}
\begin{table}[!t]
\begin{center}
\begin{tabular}{|c||c|l|}\hline
{\sc Fields}&{\sc Masses Squared}&{\sc Eigenstates} \\
\hline\hline
\multicolumn{3}{|c|}{\sc The $S$ -- $\snH$ -- $\snHb$  Sector}\\
\hline
%
2 real scalars &$ m_{\widehat \theta_\nu}^2=\mP^2\xsg^2\lf2\ld^2
(\xsg^2-6)+15g^2f\rg/24f^2$&$\widehat \theta_\nu$\\
&$m_{\widehat\th+}^2=\ld^2\mP^4\xsg^2\lf1+6\ck\rg/12J^2f^3\simeq4\Hhi^2$&$
\widehat\th_{+}$\\
1 complex scalar &$ m_{\widehat S}^2=\ld^2\mP^2\xsg^2{ \lf 12 +
\xsg^2 \bar f\rg\lf6 \kx f-1\rg/6f^2\bar f}$&$\widehat S$\\
[0.5mm] \hline\hline
\multicolumn{3}{|c|}{\sc The $u_{H\ca}^c$ --
$\bar u_{H\ca}^c$ (${\rm a}=1,2,3$) and $\eH$ -- $\eHb$ Sectors}\\
\hline
$2(3+1)$ real scalars &$ m_{\widehat u-}^2=\mP^2\xsg^2\lf\ld^2
(\xsg^2-3)+3g^2f\rg/12f^2$&$\widehat u^{\rm a}_{1-},\>\widehat
u^{\rm a}_{2+},$\\
&$m^2_{\widehat e-}=m^2_{\widehat u-}$&$\widehat e_{1-},\>\widehat e_{2+}$\\
\hline\hline
\multicolumn{3}{|c|}{\sc The $d_{H\ca}^c$ --
$\bar d_{H\ca}^c$ and $g_\ca^c$ -- $\bar g_\ca^c$ (${\ca}=1,2,3$) Sectors}\\
\hline
$3\cdot 8$ real scalars &$ m_{\widehat g}^2=\mP^2\xsg^2\lf\ld^2
\xsg^2+24\lHb^2f\rg/24f^2$&$\widehat g^{\rm a}_1,\widehat g^{\rm a}_2$\\
&$ m_{\widehat{\bar g}}^2=\mP^2\xsg^2\lf\ld^2
\xsg^2+24\lH^2f\rg/24f^2$&$\widehat{\bar g}^{\rm a}_1,\widehat{\bar g}^{\rm a}_2$\\
&$ m_{\widehat d+}^2=\mP^2\xsg^2\lf\ld^2
+4\lH^2f\rg/4f^2$&$\widehat{d}^{\rm a}_{1+},\widehat{d}^{\rm a}_{2-}$\\
&$ m_{\widehat d-}^2=\mP^2\xsg^2\lf\ld^2
\lf\xsg^2-3\rg+12\lH^2f\rg/12f^2$&$\widehat{d}^{\rm
a}_{1-},\widehat{d}^{\rm a}_{2+}$\\[.8mm] \hline
\end{tabular}\end{center}
\caption{\sl\small The scalar mass spectrum of our model along the
inflationary trajectory of \Eref{inftr}. To avoid very lengthy
formulas we neglect terms proportional to $\mpq^2$ and we assume
$\lH\simeq\lHb$ for the derivation of the masses of the scalars in
the superfields $\dH$ and $\dHb$. }\label{tab2}
\end{table}

The $8$ Goldstone bosons, associated with the modes $\what x_{1+}$
and $\what x_{2-}$ with $x=u^\ca$ and $e$, are not exactly
massless since $\what V_{{\rm HI},h}\neq0$ -- contrary to the
situation of \cref{jean} where the direction with non vanishing
$\vev{\snH}$ minimizes the potential. These
masses turn out to be $m_{x0}=\ld\mP\xsg/2f$. On the contrary, the
angular parametrization in \Eref{hpar} assists us to isolate the
massless mode $\what\th_-$, in agreement with the analysis of
\cref{linde1}.  Employing the well-known Coleman-Weinberg formula
\cite{cw}, we can compute the one-loop radiative corrections to the potential in
our model. However, these have no significant effect on the
inflationary dynamics and predictions, since the slope of the
inflationary path is generated at the classical level -- see the
expressions for $\widehat\epsilon$ and $\widehat\eta$ below.

\subsection{The Inflationary Observables}\label{fhi2}

Based on the potential of \Eref{Vhi} and keeping in mind that the
EF canonically inflaton $\what h$ is related to $h$ via
\Eref{VJe}, we can proceed to the analysis of \FHI in the EF,
employing the standard slow-roll approximation. Namely, a stage of
slow-roll nMHI is determined by the condition -- see e.g.
\cref{review, lectures}:
$$ {\ftn\sf
max}\{\widehat\epsilon(\sg),|\widehat\eta(\sg)|\}\leq 1,$$ where
\beqs\beq  \label{sr1}\widehat\epsilon=
{\mP^2\over2}\left(\frac{\Ve_{{\rm HI},\se}}{\Ve_{\rm
HI}}\right)^2={\mP^2\over2J^2}\left(\frac{\Ve_{{\rm
HI},h}}{\Ve_{\rm HI}} \right)^2\simeq
{4f_0^2\mP^4\over3\ck^2\sg^4}\eeq and \beq\label{sr2}\widehat\eta=
m^2_{\rm P}~\frac{\Ve_{{\rm HI},\se\se}}{\Ve_{\rm HI}}={\mP^2\over
J^2}\left( \frac{\Ve_{{\rm HI},hh}}{\Ve_{\rm HI}}-\frac{\Ve_{{\rm
HI},h}}{\Ve_{\rm HI}}{J_{,\sg}\over
J}\right)\simeq-{4f_0\mP^2\over3\ck\sg^2}~, \eeq\eeqs
are the slow-roll parameters and
$f_0=f\lf\vev{h}=2\Mpq\rg=1+4\ck\mpq^2$ -- see \Sref{lept}. Here
we employ \Eref{Vhio} and the following approximate relations:
\beq \label{help} J\simeq \sqrt{6}{\mP\over\sg},\>\>\>\widehat
V_{{\rm HI},\sg}\simeq {4\Vhi\over \ck h^3}f_0\mP^2
\>\>\>\mbox{and}\>\>\>\widehat V_{{\rm
HI},\sg\sg}\simeq-{12\Vhi\over \ck h^4}f_0\mP^2.\eeq
The numerical computation reveals that \FHI terminates due to the
violation of the $\widehat\epsilon$ criterion at a value of $\sg$
equal to $\sgf$, which is calculated to be
\beq \widehat\epsilon\lf\sgf\rg=1\>\Rightarrow\>
\sgf=\lf{4/3}\rg^{1/4}\mP\sqrt{f_0/\ck}. \label{sgap}\eeq

The number of e-foldings, $\widehat N_{*}$, that the scale
$k_{*}=0.002/{\rm Mpc}$ suffers during nMHI can be calculated
through the relation:
\begin{equation}
\label{Nhi}  \widehat{N}_{*}=\:\frac{1}{m^2_{\rm P}}\;
\int_{\se_{\rm f}}^{\se_{*}}\, d\se\: \frac{\Vhi}{\Ve_{{\rm
HI},\se}}= {1\over\mP^2}\int_{h_{\rm f}}^{h_{*}}\, d\sg\:
J^2\frac{\Ve_{\rm HI}}{\Ve_{{\rm HI},h}},
\end{equation}
where $h_*~[\se_*]$ is the value of $\sg~[\se]$ when $k_*$ crosses
the inflationary horizon. Given that $\sgf\ll\sg_*$, we can write
$\sg_*$ as a function of $\widehat{N}_{*}$ as follows
\beq \label{s*}
\widehat{N}_{*}\simeq{3\ck\over4f_0}{\sg_*^2-\sgf^2\over
\mP^2}\>\Rightarrow\>\sg_*=2\mP\sqrt{\widehat
N_*f_0/3\ck}\cdot\eeq

The power spectrum $\Delta^2_{\cal R}$ of the curvature
perturbations generated by $h$ at the pivot scale $k_{*}$ is
estimated as follows
\begin{equation}  \label{Prob}
\Delta_{\cal R}=\: \frac{1}{2\sqrt{3}\, \pi\mP^3} \;
\frac{\Ve_{\rm HI}(\sex)^{3/2}}{|\Ve_{{\rm
HI},\se}(\sex)|}\simeq{\ld\sg_*^2\over16\sqrt{2}\pi
f_0\mP^2}\simeq {\ld\widehat{N}_{*}\over12\sqrt{2}\pi\ck}\cdot\eeq
Since the scalars listed in \Tref{tab2} are massive enough during
nMHI, $\Delta_{\cal R}$ can be identified with its central
observational value -- see \Sref{cont} -- with almost constant
$\Ne_*$. The resulting relation reveals that $\ld$ is to be
proportional to $\ck$. Indeed we find
\beq \ld\simeq{8.4\cdot10^{-4}\pi\ck/\Ne_*}\>\Rightarrow\>
\ck\simeq20925\ld\>\>\>\mbox{for}\>\>\>\Ne_*\simeq55.\label{lan}\eeq

The (scalar) spectral index $n_{\rm s}$, its running $a_{\rm s}$,
and the scalar-to-tensor ratio $r$ can be estimated through the
relations:
\beqs\baq \label{ns} && n_{\rm s}=\: 1-6\widehat\epsilon_*\ +\
2\widehat\eta_*\simeq1-{2/\widehat N_*}, \>\>\> \\
&& \label{as} \alpha_{\rm s}
=\:{2\over3}\left(4\widehat\eta_*^2-(n_{\rm
s}-1)^2\right)-2\widehat\xi_*\simeq-2\widehat\xi_*\simeq{-2/\widehat
N^2_*}\eaq and \beq \label{rt}
r=16\widehat\epsilon_*\simeq{12/\widehat N^2_*}, \eeq\eeqs
where $\widehat\xi=\mP^4 {\Ve_{{\rm HI},\sg} \Ve_{{\rm
HI},\se\se\se}/\Vhi^2}=
\mP\,\sqrt{2\widehat\epsilon}\,\widehat\eta_{,h}/
J+2\widehat\eta\widehat\epsilon$. The variables with subscript $*$
are evaluated at $h=h_{*}$ and \eqs{sr1}{sr2} have been employed.

\section{Non-Thermal Leptogenesis}\label{pfhi}

In this section, we specify how the SUSY inflationary scenario
makes a transition to the radiation dominated era (\Sref{lept})
and give an explanation of the origin of the observed BAU
(\Sref{lept1}) consistently with the $\Gr$ constraint and the low
energy neutrino data (\Sref{lept2}).

\subsection{The Inflaton's Decay}\label{lept}

When \FHI is over, the inflaton continues to roll down towards the
SUSY vacuum, \Eref{vevs}. There is a brief stage of tachyonic
preheating \cite{preheating} which does not lead to significant
particle production \cite{garcia}.  Soon after, the inflaton
settles into a phase of damped oscillations initially around zero
-- where $\Vhio$ has a maximum -- and then around one of the
minima of $\Vhio$. Whenever the inflaton passes through zero,
particle production may occur creating mostly superheavy bosons
via the mechanism of instant preheating \cite{instant}. This
process becomes more efficient as $\ld$ decreases, and further
numerical investigation is required in order to check the
viability of the non-thermal leptogenesis scenario for small
values of $\ld$. For this reason, we restrict to $\ld$'s larger
than $0.001$, which ensures a less frequent passage of the
inflaton through zero, weakening thereby the effects from instant
preheating and other parametric resonance effects -- see Appendix
B of \cref{nmH}. Intuitively the reason is that larger $\ld$'s
require larger $\ck$'s, see \Eref{lan}, diminishing therefore
$h_{\rm f}$ given by \Eref{s*}, which sets the amplitude of the
very first oscillations.

Nonetheless the standard perturbative approach to the inflaton
decay provides a very efficient decay rate. Namely, at the SUSY
vacuum $\snH$ and $\snHb$ acquire the v.e.vs shown in \Eref{vevs}
giving rise to the masses of the (canonically normalized) inflaton
$\what{\delta h}=\lf h-2\Mpq\rg/J_0$ and RH neutrinos,
$\what\nu_i^c$,  which are given, respectively, by
\beq \label{masses} \mbox{\small\sf(a)}\>\>\>
\msn=\sqrt{2}{\ld\Mpq\over\vev{J}f_0}\>\>\>\mbox{and}\>\>\>\mbox{\small\sf
(b)}\>\>\>\mrh[i]=2{\ld_{i\nu^c}\Mpq^2\over\Ms\sqrt{f_0}},\eeq
where $f_0$ is defined below \Eref{sr2} and $\bar f_0=
f_0+24\ck^2\mpq^2\simeq J_0^2$. Here, we assume the existence of a
term similar to the second one inside $\ln$ of \Eref{Kcom} for
$\nu_i^c$ too.

For larger $\ld$'s $\vev{J}=J(h=2\Mpq)$ ranges from $3$ to $90$
and so $\msn$ is kept independent of $\ld$ and almost constant at
the level of $10^{13}~\GeV$. Indeed, if we express $\what{\delta
h}$ as a function of ${\delta h}$ through the relation
\beq \label{Jo} {\what{\delta h}\over{\delta h}}\simeq
J_0\>\>\>\mbox{where}\>\>\>
J_0=\sqrt{1+{3\over2}\mP^2f^2_{,\sg}\lf\vev{h}\rg}=\sqrt{1+24\ck^2\mpq^2}\eeq
%
we find
\beq \label{mqa} \msn\simeq{\sqrt{2}\ld\Mpq\over
f_0J_0}\simeq{\ld\mP\over2\sqrt{3}\ck}\simeq{10^{-4}\mP\over4.2\sqrt{3}}\simeq3\cdot10^{13}~\GeV
\>\>\>\mbox{for}\>\>\>\ld\gtrsim{10^{-4}\over4.2\sqrt{6}\mpq}\simeq1.3\cdot10^{-3}\eeq
where we make use of \Eref{lan} -- note that $f_0\simeq1$. The
derivation of the (s)particle spectrum, listed in Table 2, at the SUSY vaccum of the
model reveals \cite{nmH} that perturbative decays of $\what{\delta
h}$ into these massive particles are kinematically forbidden and
therefore, narrow parametric resonance \cite{preheating} effects
are absent. Also $\what{\delta h}$ can not decay via
renormalizable interaction terms to SM particles.

The inflaton can decay into a pair of $\what\nu^c_{i}$'s through
the following lagrangian terms:
\beq \label{l1}{\cal L}_{{\rm I}\nu_i^c} =
-\ld_{i\nu^c}{\Mpq\over\Ms}{f_0\over
J_0}\lf1-12\ck\mpq^2\rg\what{\delta
h}\what\nu^c_i\what\nu^c_i+{\rm h.c.}\,. \eeq
From \Eref{l1} we deduce that the decay of $\what{\delta h}$ into
$\what \nu^c_i$ is induced by two lagrangian terms. The first one
originates exclusively from the non-renormalizable term of
\Eref{Whi} -- as in the case of a similar model in \cref{jean}.
The second term is a higher order decay channel due to the SUGRA
lagrangian -- cf. \cref{Idecay}. The interaction in \Eref{l1}
gives rise to the following decay width
\beq \GNsn[j]={c_{{\rm
I}j\what\nu^c}^2\over64\pi}\msn\sqrt{1-{4\mrh[
j]^2\over\msn^2}}\>\>\>\mbox{with}\>\>\> c_{{\rm
I}j\what\nu^c}={\mrh[j]\over\Mpq}{f_0^{3/2}\over
J_0}\lf1-12\ck\mpq^2\rg, \label{Gpq}\eeq
where $\mrh[j]$ is the Majorana mass of the $\wrhn[j]$'s into
which the inflaton can decay. The implementation -- see
\Sref{lept2} -- of the seesaw mechanism for the derivation of the
light-neutrinos masses, in conjunction with the $\Ggut$ prediction
$\mD[3]\simeq m_t$ and our assumption that $\mD[1]<\mD[2]\ll
\mD[3]$ -- see \Sref{cont1} -- results to $2\mrh[3]>\msn$.
Therefore, the kinematically allowed decay channels of
$\what{\delta h}$ are those into $\what\nu^c_j$ with $j=1$ and 2.
Note that the decay of the inflaton to the heaviest of the
$\wrhn[j]$'s ($\what\nu^c_3$) is also disfavored by the $\Gr$
constraint -- see below.

In addition, there are SUGRA-induced \cite{Idecay} -- i.e., even
without direct superpotential couplings -- decay channels of the
inflaton to the MSSM particles via non-renormalizable interaction
terms. For a typical trilinear superpotential term of the form
$W_y=yXYZ$, we obtain the effective interactions described by the
langrangian part
\beq \label{l2} {\cal L}_{{{\rm I}}y} =
6y\ck{\Mpq\over\mP^2}{f_0^{3/2}\over 2J_0}\what{\delta h}\lf \what
X\what\psi_{Y}\what\psi_{Z}+\what
Y\what\psi_{X}\what\psi_{Z}+\what
Z\what\psi_{X}\what\psi_{Y}\rg+{\rm h.c.}\,, \eeq
where $y$ is a Yukawa coupling constant and $\psi_X, \psi_{Y}$ and
$\psi_{Z}$ are the chiral fermions associated with the superfields
$X, Y$ and $Z$. Their scalar components are denoted with the
superfield symbol.
Taking into account the terms of \Eref{wmssm}
and the fact that the adopted SUSY GUT predicts YU for the 3rd
generation at $\Mpq$, we conclude that the interaction above gives
rise to the following 3-body decay width
\beq \Ghsn={14 c_{{{\rm
I}}y}^2\over512\pi^3}\msn^3\simeq{3y_{33}^2
\over64\pi^3}f_0^3\lf\msn\over\mP\rg^2\msn\>\>\>\mbox{where}\>\>\>
c_{{{\rm I}}y}=6y_{33}\ck{\Mpq\over\mP^2}{f_0^{3/2}\over J_0},
\label{Gpq1}\eeq
with $y_{33}\simeq(0.55-0.7)$ being the common Yukawa coupling
constant of the third generation computed at the $\msn$ scale, and
summation is taken over color, weak and hypercharge degrees of
freedom, in conjunction with the assumption that $\msn<2\mrh[3]$.

Since the decay width of the produced $\what\nu^c_{j}$ is much
larger than \Gsn -- see below -- the reheating temperature, $\Trh$,
is exclusively determined by the inflaton decay and is given by
\cite{quin}
\beq \label{T1rh} \Trh=
\left(72\over5\pi^2g_{*}\right)^{1/4}\sqrt{\Gsn\mP}
\>\>\>\mbox{with}\>\>\>\Gsn=\GNsn[1]+\GNsn[2]+\Ghsn,\eeq
where $g_{*}$ counts the effective number of relativistic degrees
of freedom at temperature $\Trh$. For the MSSM spectrum plus the
particle content of the superfields $P$ and $\bar P$ we find
$g_{*}\simeq228.75+4(1+7/8)=236.25$.

\subsection{Lepton-Number and Gravitino Abundances}\label{lept1}

If $\Trh\ll\mrh[i]$, the out-of-equilibrium condition \cite{baryo}
for the implementation of nTL is automatically satisfied.
Subsequently $\what\nu^c_{i}$ decay into $H_u$ and $L_i^*$ via the
tree-level couplings derived from the second term in the RHS of
Eq.~(\ref{wmssm}). Interference between tree-level and one-loop
diagrams generates a lepton-number asymmetry (per $\what\nu^c_j$
decay) $\ve_j$ \cite{baryo}, when CP conservation is violated. The
resulting lepton-number asymmetry after reheating can be partially
converted through sphaleron effects into baryon-number asymmetry.
In particular, the $B$ yield can be computed as
\beq {\small\sf
(a)}\>\>\>Y_B=-0.35Y_L\>\>\>\mbox{with}\>\>\>{\small\sf
(b)}\>\>\>Y_L=2{5\over4}
{\Trh\over\msn}\sum_{j=1}^2{\GNsn[j]\over\Gsn}\ve_j\cdot\label{Yb}\eeq
The numerical factor in the RHS of \sEref{Yb}{a} comes from the
sphaleron effects, whereas the one ($5/4$) in the RHS of
\sEref{Yb}{b} is due to the slightly different calculation
\cite{quin} of $\Trh$ -- cf.~\cref{baryo1}. In the major part of
our allowed parameter space -- see \Sref{num} -- $\Gsn\simeq\Ghsn$
and so the involved branching ratio of the produced $\what\nu^c_i$
is given by
\beq
{\GNsn[1]+\GNsn[2]\over\Gsn}\simeq{\GNsn[2]\over\Ghsn}={\pi^2\lf1
-12\ck\mpq^2\rg^2\over72\ck^2y_{33}^2\mpq^4}{\mrh[2]^2\over\msn^2}\cdot\label{Gmqq}\eeq
For $\mrh[2]\simeq \lf10^{11}-10^{12}\rg~\GeV$ the ratio above
takes adequately large values so that $Y_L$ is sizable. Therefore,
the presence of more than one inflaton decay channels does not
invalidate the scenario of nTL.

It is worth emphasizing, however, that if $M_{1\nu^c}\lesssim
10\Trh$, part of the $Y_L$ can be washed out due to $\what\nu^c_1$
mediated inverse decays and $\Delta L=1$ scatterings -- this
possibility is analyzed in \cref{senoguz}. Trying to avoid the
relevant computational complications we limit ourselves to cases
with $\mrh[1]\gtrsim10\Trh$, so as any washout of the
non-thermally produced $Y_L$ is evaded. On the other hand, $Y_L$
is not erased by the $\Delta L=2$ scattering processes
\cite{erasure} at all temperatures $T$ with $100~\GeV\lesssim
T\lesssim\Trh$ since $Y_L$ is automatically protected by SUSY
\cite{ibanez} for $10^7~\GeV\lesssim T\lesssim\Trh$ and for
$T\lesssim10^7~\GeV$ these processes are well out of equilibrium
provided that that mass of the heaviest light neutrino is
$10~\eV$. This constraint, however, is overshadowed by a more
stringent one induced by WMAP7 data \cite{wmap} -- see
\Sref{cont1}.

The required for successful nTL $\Trh$ must be compatible with
constraints on the $\Gr$ abundance, $Y_{\Gr}$, at the onset of
\emph{nucleosynthesis} (BBN). This is estimated to be
\cite{kohri}:
\beq\label{Ygr} Y_{\Gr}\simeq c_{\Gr}
\Trh\>\>\>\mbox{with}\>\>\>c_{\Gr}=1.9\cdot10^{-22}/\GeV,\eeq
where we assume that $\Gr$ is much heavier than the gauginos. Let
us note that non-thermal $\Gr$ production within SUGRA is
\cite{Idecay} also possible. However, we here prefer to adopt the
conservative approach based on the estimation of $Y_{\Gr}$ via
\Eref{Ygr} since the latter $\Gr$  production depends on the
mechanism of SUSY breaking.

Both \eqs{Yb}{Ygr} yield the correct values of the $B$ and $\Gr$
abundances provided that no entropy production occurs for
$T<\Trh$. This fact can be easily achieved within our setting. The
mass spectrum of the $P$-$\bar P$ system is comprised by axion and
saxion $P_-=(\bar P-P)/\sqrt{2}$, axino $\psi_-=(\psi_{\bar
P}-\psi_P)/\sqrt{2}$, a higgs, $P_+=(\bar P+P)/\sqrt{2}$, and a
higgsino, $\psi_+=(\psi_{\bar P}+\psi_P)/\sqrt{2}$, with mass of
order $1~\TeV$ and $\psi$ denoting a Weyl spinor. The higgs and
higgsinos can decay to lighter higgs and higgsinos before
domination \cite{curvaton}. Regarding the saxion, $P_{-}$, we can
assume that its decay mode to axions is suppressed (w.r.t the ones
to gluons, higgses and higgsinos \cite{Baer, senami}) and the
initial amplitude of its oscillations is equal to
$f_a\simeq10^{12}~\GeV$. Under these circumstances, it can
\cite{Baer} decay before domination too, and evades \cite{senami}
the constraints from the effective number of neutrinos for the
$f_a$'s and $\Trh$'s encountered in our model. As a consequence of
its relatively large decay temperature, the LSPs produced by the
saxion decay are likely to be thermalized and therefore, no upper
bound on the saxion abundance is \cite{senami} to be imposed.
Finally, axino can not play the role of LSP due to its large
expected mass and the relatively high $\Trh$'s encountered in our
set-up which result to a large \emph{Cold Dark Matter} (CDM)
abundance. Nonetheless, it may enhance non-thermally the abundance
of a higgsino-like neutralino-LSP, rendering it a successful CDM
candidate.

\subsection{Lepton-Number Asymmetry and Neutrino Masses}\label{lept2}

As mentioned above, the decay of $\what\nu^c_2$ and
$\what\nu^c_1$, emerging from the $\what{\delta h}$ decay, can
generate a lepton asymmetry, $\ve_i$ (with $i=1,2$) caused by the
interference between the tree and one-loop decay diagrams,
provided that a CP-violation occurs in $h_{Nij}$'s. The produced
$\ve_i$ can be expressed in terms of the Dirac mass matrix of
$\nu_i$, $m_{\rm D}$, defined in a basis (called
$\sni$-\emph{basis} henceforth) where $\sni$ are mass eigenstates,
as follows:
\beqs\beq\ve_i =\sum_{i\neq j}{
\im\left[(\mD[]^\dag\mD[])_{ij}^2\right]\over8\pi\vev{H_u}^2(\mD[]^\dag\mD[])_{ii}}
\bigg( F_{\rm S}\lf x_{ij},y_i,y_j\rg+F_{\rm V}(x_{ij})\bigg) ,
\label{el}\eeq where we take $\vev{H_u}\simeq174~\GeV$, for large
$\tan\beta$ and \beq
x_{ij}:={\mrh[j]\over\mrh[i]}\>\>\>\mbox{and}\>\>\>y_i:={\Gm[i\nu^c]\over\mrh[i]}=
{(\mD[]^\dag\mD[])_{ii}\over8\pi\vev{H_u}^2}\eeq (with
$i,j=1,2,3$). Also $F_{\rm V}$ and $F_{\rm S}$ represent,
respectively, the contributions from vertex and self-energy
diagrams which in SUSY theories read \cite{resonant1} \beq F_{\rm
V}\lf x\rg=-x\ln\lf1+ x^{-2}\rg~~\mbox{and}~~F_{\rm S}\lf
x,y,z\rg={-2x(x^2-1)\over\lf x^2-1\rg^2+\lf x^2z-y\rg^2}\cdot\eeq
\eeqs Note that for strongly hierarchical $\mrh[]$'s  with
$x_{ij}\gg1$ and $x_{ij}\gg y_i,y_j$, we obtain  the well-known
approximate result \cite{frigerio, senoguz} \beq \label{hieF}
F_{\rm V}+F_{\rm S}\simeq-3/x_{ij}^2.\eeq

The involved in \Eref{el} $\mD[]$ can be diagonalized if we define
a basis -- called \emph{weak basis} henceforth -- in which the
lepton Yukawa couplings and the $SU(2)_{\rm L}$ interactions are
diagonal in the space of generations. In particular we have
\beq \label{dD} U^\dag\mD[]U^{c\dag}=d_{\rm
D}=\diag\lf\mD[1],\mD[2], \mD[3]\rg,\eeq where $U$ and $U^c$ are
$3\times3$ unitary matrices which relate $L_i$ and $\sni$ (in the
$\sni$-basis) with the ones $L'_i$ and $\nu^{c\prime}_i$ in the
weak basis as follows:
\beq L'= L U\>\>\> \mbox{and}\>\>\>\nu^{c\prime}=U^c \nu^c.\eeq
Here, we write LH lepton superfields, i.e. $SU(2)_{\rm L}$ doublet
leptons, as row 3-vectors in family space and RH anti-lepton
superfields, i.e. $SU(2)_{\rm L}$ singlet anti-leptons, as column
3-vectors. Consequently, the combination $\mD[]^\dag\mD[]$
appeared in \Eref{el} turns out to be a function just of $d_{\rm
D}$ and $U^c$. Namely, \beq\mD[]^\dag\mD[]=U^{c\dag} d^\dag_{\rm
D}d_{\rm D}U^c. \label{mDD}\eeq

The connection of the nTL scenario with the low energy neutrino
data can be achieved through the seesaw formula, which gives the
light-neutrino mass matrix $m_\nu$ in terms of $\mD[i]$ and
$\mrh[i]$. Working in the $\sni$-basis, we have
\beq \label{seesaw} m_\nu= -m_{\rm D}\ d_{\nu^c}^{-1}\ m_{\rm
D}^{\tr},~~\mbox{where}~~ d_{\nu^c}=
\diag\lf\mrh[1],\mrh[2],\mrh[3]\rg \eeq with
$\mrh[1]\leq\mrh[2]\leq\mrh[3]$ real and positive. Solving
\Eref{dD} w.r.t $\mD[]$ and inserting the resulting expression in
\Eref{seesaw} we extract the mass matrix
\beq \label{bmn} \bar m_\nu=U^\dag m_\nu U^*=-d_{\rm
D}U^cd_{\nu^c}^{-1}U^{c\tr}d_{\rm D},\eeq which can be
diagonalized by the unitary PMNS matrix satisfying \beq \bar
m_\nu=U_\nu^*\ \diag\lf\mn[1],\mn[2],\mn[3]\rg\
U^\dag_\nu\label{mns1}\eeq and parameterized as follows:
\beq \label{mns2} U_\nu = \mtn{c_{12}c_{13}}{s_{12}c_{13}}{s_{13}
e^{-i\delta}} {-c_{23}s_{12}-s_{23}c_{12}s_{13}
e^{i\delta}}{c_{23}c_{12}-s_{23}s_{12}s_{13}
e^{i\delta}}{s_{23}c_{13}} {s_{23}s_{12}-c_{23}c_{12}s_{13}
e^{i\delta}}{-s_{23}c_{12}-c_{23}s_{12}s_{13}
e^{i\delta}}{c_{23}c_{13}}\cdot\mtn{e^{-i\varphi_1/2}}{0}{0}{0}{e^{-i\varphi_2/2}}{0}{0}{0}{1},
\eeq with $c_{ij}:=\cos \theta_{ij}$, $s_{ij}:=\sin \theta_{ij}$,
$\delta$ the CP-violating Dirac phase and $\varphi_1$ and
$\varphi_2$ the two CP-violating Majorana phases.

Following a bottom-up approach, along the lines of \cref{nMCI,
frigerio, senoguz}, we can find $\bar m_\nu$ via \Eref{mns1} using
as input parameters the low energy neutrino observables, the CP
violating phases and adopting the normal or inverted hierarchical
scheme of neutrino masses. Taking also $\mD[i]$ as input
parameters we can construct the complex symmetric matrix \beq
W=-d_{\rm D}^{-1}\bar m_\nu d_{\rm
D}^{-1}=U^cd_{\nu^c}U^{c\tr}\label{Wm}\eeq -- see \Eref{bmn} --
from which we can extract $d_{\nu^c}$ as follows: \beq
d_{\nu^c}^{-2}=U^{c\dag}W W^\dag U^c.\label{WW}\eeq Note that $W
W^\dag$ is a $3\times3$ complex, hermitian matrix and can be
diagonalized following the algorithm described in \cref{33m}.
Having determined the elements of $U^c$ and the $\mrh[i]$'s we can
compute $\mD[]$ through \Eref{mDD} and the $\ve_i$'s through
\Eref{el}.

\section{Constraining the Model Parameters}\label{cont}

We exhibit the constraints that we impose on our cosmological
set-up in \Sref{cont1}, and delineate the allowed parameter space
of our model in Sec.~\ref{num}.

\subsection{Imposed Constraints}\label{cont1}

The parameters of our model can be restricted once we impose the
following requirements:

\begin{itemize}

\item[{\sf 1.}] According to the inflationary paradigm, the
horizon and flatness problems of the standard Big Bang cosmology
can be successfully resolved provided that the number of
e-foldings, $\widehat N_{*}$, that the scale $k_{*}=0.002/{\rm
Mpc}$ suffers during \FHI takes a certain value, which depends on
the details of the cosmological model. Employing standard methods
\cite{nmi}, we can easily derive the required $\widehat{N}_{*}$
for our model, consistently with the fact that the $P-\bar P$
system remains subdominant during the post-inflationary era.
Namely we obtain
\begin{equation}  \label{Ntot}
\widehat{N}_{*}\simeq22.5+2\ln{V_{\rm
HI}(\sg_{*})^{1/4}\over{1~{\rm GeV}}}-{4\over 3}\ln{V_{\rm
HI}(\sg_{\rm f})^{1/4}\over{1~{\rm GeV}}}+ {1\over3}\ln {T_{\rm
rh}\over{1~{\rm GeV}}}+{1\over2}\ln{f(\sg_{\rm f})\over
f(\sg_*)}\cdot
\end{equation}

\item[{\sf 2.}] The inflationary observables derived in
\Sref{fhi2} are to be consistent with the fitting \cite{wmap} of
the WMAP7, BAO and $H_0$ data. As usual, we adopt the central
value of $\Delta_{\cal R}$, whereas we allow the remaining
quantities to vary within the 95$\%$ \emph{confidence level}
(c.l.) ranges. Namely,
\beq \label{obs1} \mbox{\small\sf (a)}~~ \Delta_{\cal
R}\simeq4.93\cdot 10^{-5},~ \mbox{\small\sf
(b)}\>~\ns=0.968\pm0.024,\>\>\mbox{\small\sf (c)}\>~-0.062\leq
a_{\rm s}\leq0.018 \>\>\mbox{and}\>\>\mbox{\small\sf
(d)}\>~r<0.24\eeq

\item[{\sf 3.}] The scale $\Mpq$ can be determined by requiring
that the v.e.vs of the Higgs fields take the values dictated by
the unification of the gauge couplings within the MSSM. As we now
recognize -- cf. \cref{nmH} -- the unification scale
$\Mgut\simeq2\cdot10^{16}~\GeV$ is to be identified with the
\emph{lowest} mass scale of the model in the SUSY vacuum,
\Eref{vevs}, in order to avoid any extra contribution to the
running of the MSSM gauge couplings, i.e.,
\beq \label{Mpqf} {\small\sf (a)}~~{g\Mpq\over
\sqrt{f_0}}=\Mgut\>\Rightarrow\>\mpq={1\over2\sqrt{2c^{\rm
max}_{\mathcal R}-\ck}}~~\mbox{with}~~{\small\sf (b)}~~ c^{\rm
max}_{\mathcal R}={g^2\mP^2\over8\Mgut^2}\eeq
The requirement $2c^{\rm max}_{\mathcal R}-\ck>0$ sets an upper
bound $\ck<2c^{\rm max}_{\mathcal R}\simeq1.8\cdot10^3$, which
however can be significantly lowered if we combine \eqs{Ntot}{Nhi}
-- see \Sref{num1}.

\item[{\sf 4.}] For the realization of \FHI, we assume that $\ck$
takes relatively large values -- see e.g. \Eref{Sni1}. This
assumption may \cite{cutoff, cutoff1} jeopardize the validity of
the classical approximation, on which the analysis of the
inflationary behavior is based. To avoid this inconsistency --
which is rather questionable \cite{cutoff3, linde1} though -- we
have to check the hierarchy between the ultraviolet cut-off,
$\Ld=\mP/\ck$, of the effective theory and the inflationary scale,
which is represented by $\Vhi(\sg_*)^{1/4}$ or, less
restrictively, by the corresponding Hubble parameter, $\widehat
H_*=\Vhi(\sg_*)^{1/2}/\sqrt{3}\mP$. In particular, the validity of
the effective theory implies \cite{cutoff,cutoff1}
\beq \label{Vl}\mbox{\small\sf (a)}\>~ \Vhi(\sg_*)^{1/4}\leq\Ld
\>\>\mbox{or}\>\>\mbox{\small\sf (b)}\>~ \widehat
H_*\leq\Ld\>\>\mbox{for}\>\>\mbox{\small\sf (c)}\>~\ck\geq1.\eeq

\item[{\sf 5.}] As discussed in \Sref{lept1}, to avoid any erasure
of the produced $Y_L$ and to ensure that the inflaton decay to
$\what\nu_2$ is kinematically allowed we have to bound $\mrh[1]$
and $\mrh[2]$ respectively as follows:
\beq\label{kin} {\sf\small (a)}\>\>\mrh[1]\gtrsim
10\Trh\>\>\mbox{and}\>\>{\sf\small
(b)}\>\>\msn\geq2\mrh[2]\>\>\Rightarrow\>\>\mrh[2]\lesssim{\ld\mP\over4\sqrt{3}\ck}\simeq1.5\cdot10^{13}~
\GeV,\eeq
where we make use of \Eref{mqa}. Recall that we impose also the
restriction $\ld\geq0.001$ which allows us to ignore effects of
instant preheating \cite{instant,nmH}.

\item[{\sf 6.}] As discussed below \Eref{wmssm}, the adopted  GUT
predicts YU at $\Mgut$. Assuming negligible running of $\mD[3]$
from $\Mgut$ until the scale of nTL, $\Lambda_L$, which is taken
to be $\Lambda_L=\msn$, we end up with the requirement:
\beq\label{mtop}\mD[3](\msn)=m_t(\msn)\simeq (100-120)~\GeV.\eeq
where $m_t$ is the top quark mass and the numerical values
correspond to $y_{33}(\msn)=(0.55-0.7)$ -- cf. \cref{fermionM} --
found \cite{shafi,lpnova1} working in the context of several MSSM
versions with $\tan\beta\simeq50$ and taking into account the SUSY
threshold corrections. As regards the lighter generation, we limit
ourselves in imposing just a mild hierarchy between $\mD[1]$ and
$\mD[2]$, i.e., $\mD[1]<\mD[2]\ll\mD[3]$ since it is not possible
to achieve a simultaneous fulfilment of all the residual
constraints if we impose relations similar to \Eref{mtop} -- cf.
\cref{frigerio, branco, senoguz}.

\item[{\sf 7.}] From the solar, atmospheric, accelerator and
reactor neutrino experiments we take into account the inputs
listed in \Tref{tabn} on the neutrino mass-squared differences
$\Delta m^2_{21}$ and $\Delta m^2_{31}$, on the mixing angles
$\theta_{ij}$ and on the CP-violating Dirac phase, $\delta$ for
normal [inverted] neutrino mass hierarchy \cite{valle12} -- see
also \cref{lisi12}. In particular, $\mn[i]$'s can be determined
via the relations:
\beq \label{mns} \mn[2]=\sqrt{\mn[1]^2+\Delta m^2_{21}}
~~\mbox{and}~~\left\{\bem
\mn[3]=\sqrt{\mn[1]^2+\Delta m^2_{31}}, \hfill& \mbox{for
\emph{normally ordered} (NO) $\mn[]$'s} \cr
\mbox{or} &\cr
\mn[1]=\sqrt{\mn[3]^2+\left|\Delta m^2_{31}\right|},\hfill   &
\mbox{for \emph{invertedly ordered} (IO) $\mn[]$'s} \cr\eem
\right.\eeq

\renewcommand{\arraystretch}{1.6}
\begin{table}[!t]
\begin{center}
\begin{tabular}{|c|c|c|}\hline
{\sc Parameter }&\multicolumn{2}{c|}{\sc Best Fit
$\pm1\sigma$}\\\cline{2-3} &{\sc Normal}&{\sc Inverted}
\\\cline{2-3} &\multicolumn{2}{c|}{\sc Hierarchy} \\
\hline\hline
$\Delta
m^2_{21}/10^{-3}\eV^2~~~~$&\multicolumn{2}{c|}{$7.62\pm0.19$}\\\cline{2-3}
$\Delta
m^2_{31}/10^{-3}\eV^2$&$~~~2.53^{+0.08}_{-0.10}~~~$&$~~~-2.4^{+0.10}_{-0.07}~~~$\\\hline
$\sin^2\theta_{12}$ &
\multicolumn{2}{c|}{$0.320^{+0.015}_{-0.017}$}\\\cline{2-3}
$\sin^2\theta_{13}$&$0.026^{+0.003}_{-0.004}$&$0.027^{+0.003}_{-0.004}$\\
$\sin^2\theta_{23}$&$0.49^{+0.08}_{-0.05}$&$0.53^{+0.05}_{-0.07}$\\\hline
$\delta/\pi~~$&$~~0.83^{+0.54}_{-0.64}~~~$&$~~~0.07~~~$\\
\hline
\end{tabular}\end{center}
\caption{\sl\small Low energy experimental neutrino data for
normal or inverted hierarchical neutrino masses. In the second
case the full range $(0-2\pi)$ is allowed at $1\sigma$ for the
phase $\delta$.}\label{tabn}
\end{table}

The sum of $\mn[i]$'s can be bounded from above by the WMAP7 data
\cite{wmap} \beq\label{sumn} \mbox{$\sum_i$}
\mn[i]\leq0.58~{\eV}\eeq at 95\% c.l. This is more restrictive
than the 95\% c.l. upper bound arising from the effective electron
neutrino mass in $\beta$-decay \cite{2beta}: \beq \label{mbeta}
m_{\beta}:=\left|\mbox{$\sum_i$}
U_{1i\nu}^2\mn[i]\right|\leq2.3~\eV.\eeq However, in the future,
the KATRIN experiment \cite{katrin} expects to reach the
sensitivity of $m_\beta\simeq0.2~\eV$ at $90\%$ c.l.

\item[{\sf 8.}] The interpretation of BAU through nTL dictates
\cite{wmap} at 95\% c.l.
%
%
\beq Y_B=\lf8.74\pm0.42\rg\cdot10^{-11}\>\>\Rightarrow\>\>
8.32\leq 10^{11}Y_B\leq9.16.\label{BAUwmap}\eeq

\item[{\sf 9.}] In order to avoid spoiling the success of the BBN,
an upper bound on $Y_{\Gr}$ is to be imposed depending on the
$\Gr$ mass, $m_{\Gr}$,  and the dominant $\Gr$ decay mode. For the
conservative case where $\Gr$ decays with a tiny hadronic
branching ratio, we have \cite{kohri}
\beq  \label{Ygw} Y_{\Gr}\lesssim\left\{\bem
10^{-14}\hfill \cr
10^{-13}\hfill \cr
10^{-12}\hfill \cr\eem
\right.\>\>\>\mbox{for}\>\>\>m_{\Gr}\simeq\left\{\bem
0.69~{\rm TeV}\hfill \cr
10.6~{\rm TeV}\hfill \cr
13.5~{\rm TeV.}\hfill \cr\eem
\right.\eeq
As we see below, this bound is achievable  within our model only
for $m_{\Gr}\gtrsim10~\TeV$. Taking into account that the soft
masses of the scalars are not necessarily equal to $m_{\Gr}$, we
do not consider such a restriction as a very severe tuning of the
SUSY parameter space. Using \Eref{Ygr} the bounds on $Y_{\Gr}$ can
be translated into bounds on $\Trh$. Specifically we take
$\Trh\simeq\lf0.53-5.3\rg\cdot10^8~\GeV$
[$\Trh\simeq\lf0.53-5.3\rg\cdot10^9~\GeV$] for
$Y_{\Gr}\simeq\lf0.1-1\rg\cdot10^{-13}$
[$Y_{\Gr}\simeq\lf0.1-1\rg\cdot10^{-12}$].

\end{itemize}

Let us, finally, comment on the  axion isocurvature perturbations
generated in our model. Indeed, since the PQ symmetry is broken
during nMHI, the axion acquires quantum fluctuations as all the
almost massless degrees of freedom. At the QCD phase transition,
these fluctuations turn into isocurvature perturbations in the
axion energy density, which means that the partial curvature
perturbation in axions is different than the one in photons. The
results of WMAP put stringent bounds on the possible CDM
isocurvature perturbation. Namely, taking into account the WMAP7,
BAO and $H_0$ data on the parameter $\alpha_0$ we find the
following bound for the amplitude of the CDM isocurvature
perturbation
\beq \label{Swmap} |{\cal S}_{\rm c}|=\Delta_{\cal
R}\sqrt{\alpha_0\over1-\alpha_0}\lesssim1.5\cdot10^{-5}\>\>\mbox{at}
~~95\%~\mbox{c.l.}\eeq
On the other, $|{\cal S}_{\rm c}|$ due to axion, can be estimated
by
\beq \label{Sc} |{\cal S}_{\rm c}|={\Omega_a\over\Omega_{\rm
c}}\,{\Hhi\over\pi|\theta_{\rm
I}|\what\phi_{P*}}~~\mbox{with}~~{\Omega_a\over\Omega_{\rm
c}}\simeq \theta^2_{\rm I} \lf f_a \over 1.56\cdot
10^{11}~\GeV\rg^{1.175}\eeq
where $\Omega_a~[\Omega_{\rm c}]$ is the axion [CDM] density
parameter, $\what\phi_{P*}\sim10^{16}~\GeV$ \cite{curvaton}
denotes the field value of the PQ scalar when the cosmological
scales exit the horizon and $\theta_{\rm I}$ is the initial
misalignment angle which lies \cite{curvaton} in the interval
$\left[-\pi/6,\pi/6\right]$. Satisfying \Eref{Swmap} requires
$|\theta_{\rm I}|\lesssim\pi/70$ which is a rather low but not
unacceptable value. Therefore, a large axion contribution to CDM
is disfavored within our model.

\subsection{Numerical Results}\label{num}

As can be seen from the relevant expressions in Secs.~\ref{fhim}
and \ref{pfhi}, our cosmological set-up depends on the parameters:
$$\ld,\>\lH,\>\lHb,\>\kx,~g,~y_{33},~\mn[\ell],~\mD[i],~\varphi_1~\mbox{and}~\varphi_2,$$ where
$\mn[\ell]$ is the low scale mass of the lightest of $\nu_i$'s and
can be identified with $\mn[1]~[\mn[3]]$ for NO [IO] neutrino mass
spectrum. Recall that we determine $\Mpq$ via \Eref{Mpqf} with
$g=0.7$.  We do not consider $\ck$ and $\ld_{i\nu^c}$ as
independent parameters since $\ck$ is related to $m$ via
\Eref{lan} while $\ld_{i\nu^c}$ can be derived from the last six
parameters above which affect exclusively the $Y_L$ calculation
and can be constrained through the requirements 5 - 9 of
\Sref{cont1}. Note that the $\ld_{i\nu^c}$'s can be replaced by
$\mrh[i]$'s given in \sEref{masses}{b} keeping in mind that
perturbativity requires $\ld_{i\nu^c}\leq\sqrt{4\pi}$ or
$\mrh[i]\leq10^{16}~\GeV$. Note that if we replace $\Ms$ with
$\mP$ in \Eref{Whi}, we obtain a tighter bound, i.e.,
$\mrh[i]\leq2.3\cdot10^{15}~\GeV$. Our results are essentially
independent of $\lH,\>\lHb$ and $\kx$, provided that we choose
some relatively large values for these so as $m^2_{\what u-},
m^2_{\what d-}$ and $m_{\what S}^2$ in \Tref{tab2} are positive
for $\ld<1$. We therefore set $\lH=\lHb=0.5$ and $\kx=1$
throughout our calculation. Finally $\Trh$ can be calculated
self-consistently in our model as a function of $\msn$,
$\mrh[2]\gg\mrh[1]$ and the unified Yukawa coupling constant
$y_{33}$ -- see \Sref{lept} -- for which we take $y_{33}=0.6$.

Summarizing, we set throughout our calculation:
\beq \kx=1,~\lH=\lHb=0.5,g=0.7\>\>\mbox{and}\>\>y_{33}=0.6.\eeq
The selected values for the above quantities give us a wide and
natural allowed region for the remaining fundamental parameters of
our model, as we show below concentrating separately in the
inflationary period (\Sref{num1}) and in the stage of nTL
(\Sref{num2}).

\subsubsection{The Inflationary Stage}\label{num1}

\begin{figure}[!t]\vspace*{-.15in}
\hspace*{-.3in}
\begin{minipage}{8in}
\includegraphics[height=2.3in]{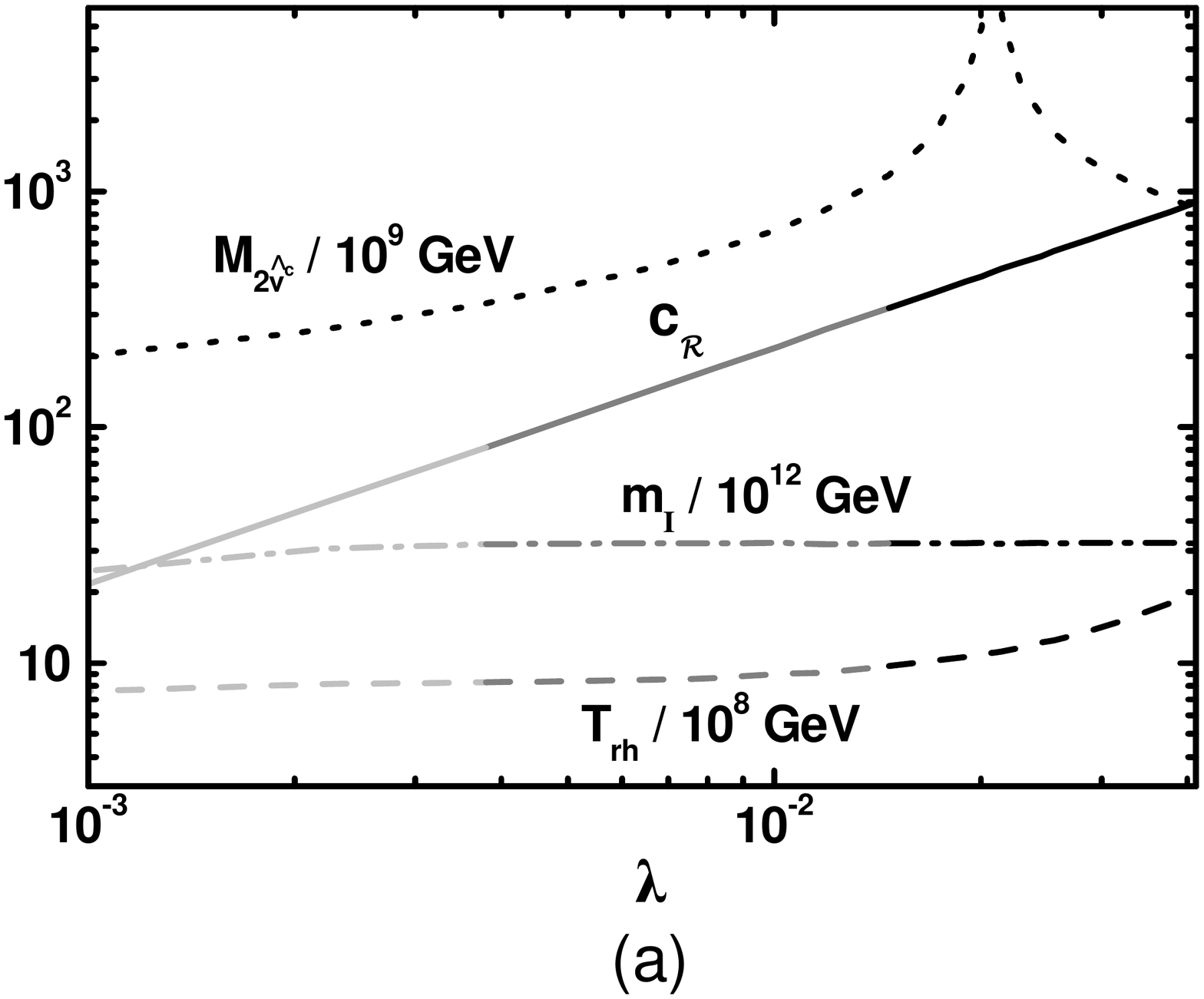}\hspace*{-0.8cm}
\includegraphics[height=2.3in]{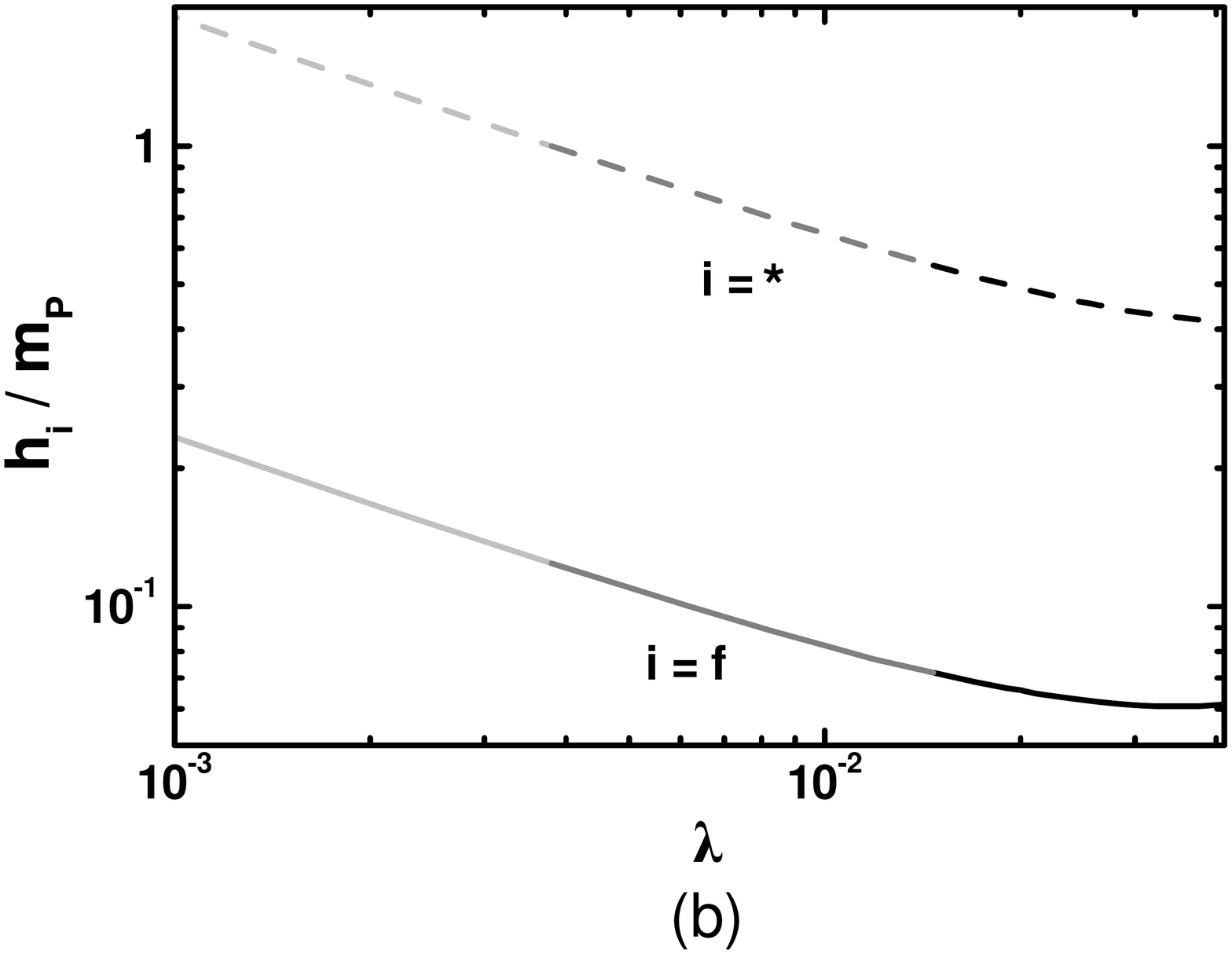}\hfill
\end{minipage}\vspace*{-0.5cm}
\hfill \caption{\sl\small  The allowed (by all the imposed
constraints) values of $\ck$ (solid line), $\Trh$ -- given by
\Eref{T1rh} -- (dashed line), $\msn$ (dot-dashed line) and
$\mrh[2]$ (dotted line) [$\sg_{\rm f}$ (solid line) and $\sg_*$
(dashed line)] versus $\ld$  (a) [(b)] for $\kx=1$, $\lH=\lHb=0.5$
and $y_{33}=0.6$. The light gray and gray segments denote values
of the various quantities satisfying \sEref{Vl}{a} too, whereas
along the light gray segments we obtain
$\sg_*\geq\mP$.}\label{fig1}
\end{figure}


In this part of our numerical code, we use as input parameters
$\sg_*,\mD[{\rm 2}]\gg\mD[1]$ and $\ck$. For every chosen
$\ck\geq1$ and $\mD[{\rm 2}]$, we restrict $\ld$ and $\sg_*$ so
that the conditions \Eref{Ntot} and (\ref{obs1}{\sf a}) are
satisfied. In our numerical calculations, we use the complete
formulas for the slow-roll parameters and $\Delta_{\cal R}$ in
\eqss{sr1}{sr2}{Prob} and not the approximate relations listed in
\Sref{fhi2} for the sake of presentation. Our results are
displayed in \Fref{fig1}, where we draw the allowed values of
$\ck$ (solid line), $\Trh$ (dashed line), the inflaton mass,
$\msn$ (dot-dashed line) and $\mrh[2]$ (dotted line) -- see
\Sref{lept} -- [$\sg_{\rm f}$ (solid line) and $\sg_*$ (dashed
line)] versus $\ld$ (a) [(b)] for the $\mD[{\rm 2}]$'s required
from \Eref{BAUwmap} and for the parameters adopted along the black
dashed line of \Fref{fig2} -- see \Sref{num2}. The required via
Eq.~(\ref{Ntot}) $\Ne_*$ remains almost constant and close to
$54.5$.

The lower bound of the depicted lines comes from the saturation of
the \sEref{Vl}{c}. The constraint of \sEref{Vl}{b} is satisfied
along the various curves whereas \sEref{Vl}{a} is valid only along
the gray and light gray segments of these. Along the light gray
segments, though, we obtain $\sg_*\geq\mP$. The latter regions of
parameter space are not necessarily excluded, since the energy
density of the inflaton remains sub-Planckian and so, corrections
from quantum gravity can still be assumed to be small.  As $\ck$
increases beyond $906$, $f_0$ becomes much larger than $1$, $\what
N_*$ derived by \Eref{Nhi} starts decreasing and therefore, nMHI
fails to fulfil \Eref{Ntot}. This can be understood by the
observation that $\what N_*$, approximated fairly by \Eref{s*},
becomes monotonically decreasing function of $\ck$ for $\ck>c^{\rm
max}_{\mathcal R}$ where $c^{\rm max}_{\mathcal R}$ can be found
by the condition
\beq {d\what N_*\over d\ck}\simeq{3h_*^2\over4\mP^2}{\lf c^{\rm
max}_{\mathcal R}-\ck\rg\over c^{\rm max}_{\mathcal
R}}=0~\Rightarrow~\ck\simeq c^{\rm max}_{\mathcal
R},\label{ckmax}\eeq
where $c^{\rm max}_{\mathcal R}$ is defined in \sEref{Mpqf}{b} and
\sEref{Mpqf}{a} is also taken into account. As a consequence, the
embedding of nMHI in a SUSY GUT provides us with a clear upper
bound of $\ck$. All in all, we obtain
\beq\label{res1} 0.001\lesssim
\ld\lesssim0.042\>\>\mbox{and}\>\>1\lesssim
\ck\lesssim907\>\>\mbox{for}\>\> \Ne_*\simeq54.5\eeq
When $\ck$ ranges within its allowed region, we take
$\Mpq\simeq\lf2.87-4\rg\cdot10^{16}~\GeV$.

From \sFref{fig1}{a}, we can verify our analytical estimation in
\Eref{lan} according to which $\ld$ is proportional to $\ck$. On
the other hand, the variation of $\sg_{\rm f}$ and $\sg_*$ as a
function of $\ck$ -- drawn in \sFref{fig1}{b} -- is consistent
with \eqs{sgap}{s*}. Letting $\ld$ or $\ck$ vary within its
allowed region in \Eref{res1}, we obtain
\beq\label{res} \ns\simeq0.964,\>\>
{\as}\simeq-6.35\cdot10^{-4}\>\>\mbox{and}\>\>
{r}\simeq3.6\cdot10^{-3}.\eeq
Clearly, the predicted $\as$ and $r$ lie within the allowed ranges
given in \sEref{obs1}{b} and \sEref{obs1}{c} respectively, whereas
$\ns$ turns out to be impressively close to its central
observationally favored value -- see \sEref{obs1}{a} and cf.
\cref{linde2}.

From \sFref{fig1}{a} we can conclude that $\msn$ is kept
independent of $\ld$ and almost constant at the level of
$10^{13}~\GeV$, as anticipated in \Eref{mqa}. From the same plot
we also remark that for $\ld\lesssim0.03$, $\Trh$ remains almost
constant since $\Ghsn$ dominates over $\GNsn[2]$ and
$f_0^3\simeq1$ -- see \Eref{Gpq1}. For $\ld\gtrsim0.03$,
$f_0^3\simeq1+12\ck\mpq^2$ starts to deviate from unity and so,
$\Trh$ increases with $\ck$ or $\ld$ as shown in \Fref{fig1}. The
required by \Eref{BAUwmap} $\mrh[2]$ follows the behavior of the
required $\mD[2]$ -- see \sFref{fig2}{a} of \Sref{num2}.

\subsubsection{The Stage of non-Thermal Leptogenesis}\label{num2}

\begin{table}[!t]\bec
\begin{tabular}{|c||c|c||c|c|c||c|c|}\hline
{\sc Parameters} &  \multicolumn{7}{c|}{\sc Cases}\\\cline{2-8}
&A&B& C & D& E & F&G\\ \cline{2-8} &\multicolumn{2}{c||}{\sc
Normal} & \multicolumn{3}{c||}{\sc Degenerate}&
\multicolumn{2}{c|}{\sc Inverted}
\\& \multicolumn{2}{c||}{\sc Hierarchy}&\multicolumn{3}{c||}{\sc Masses}&
\multicolumn{2}{c|}{\sc Hierarchy}\\ \hline
\multicolumn{8}{|c|}{\sc Low Scale Parameters}\\\hline
$\mn[1]/0.1~\eV$&$0.01$&$0.1$&$0.5$ & $1.$& $0.7$ & $0.5$&$0.49$\\
$\mn[2]/0.1~\eV$&$0.088$&$0.13$&$0.5$ & $1.$& $0.7$ & $0.51$&$0.5$\\
$\mn[3]/0.1~\eV$&$0.5$&$0.5$&$0.71$ & $1.1$&$0.5$ &
$0.1$&$0.05$\\\hline
$\sum_i\mn[i]/0.1~\eV$&$0.6$&$0.74$&$1.7$ & $3.1$&$1.9$ &
$1.1$&$1$\\
$m_\beta/0.1~\eV$&$0.03$&$0.013$&$0.14$ & $0.68$&$0.45$ &
$0.33$&$0.4$\\\hline
$\varphi_1$&$\pi/2$&$\pi/2$&$\pi/2$ & $-\pi/2$&$-\pi/2$ & $-\pi/2$&$-\pi/4$\\
$\varphi_2$&$0$&$-\pi/2$ &$-\pi/2$& $\pi$&$\pi$ &
$\pi$&$0$\\\hline
\multicolumn{8}{|c|}{\sc Leptogenesis-Scale Parameters}\\\hline
$\mD[1]/\GeV$&$0.4$&$0.3$&$0.8$ & $1$&$0.9$ & $0.9$&$0.9$\\
$\mD[2]/\GeV$&$9.2$&$3$&$6.5$ & $8.6$&$3.95$ & $6.6$&$9.2$\\
$\mD[3]/\GeV$&$120$&$100$&$100$ & $120$&$120$ &
$110$&$110$\\\hline
$\mrh[1]/10^{10}~\GeV$&$4.4$&$4.7$&$3.6$ & $1.2$&$1.5$ & $1.6$&$1.7$\\
$\mrh[2]/10^{12}~\GeV$&$2.7$&$0.6$&$0.65$ & $1.5$&$0.9$ & $1.5$&$2.8$\\
$\mrh[3]/10^{14}~\GeV$&$27$&$0.28$&$0.46$ & $0.4$&$0.4$ &
$3.8$&$10$\\\hline
\multicolumn{8}{|c|}{\sc Resulting $B$-Yield }\\\hline
$10^{11}Y^0_B$&$8.75$&$8.9$& $8.6$& $8.63$&$8.9$ & $9.$&$8.76$\\
$10^{11}Y_B$&$9.3$&$8.7$& $8.5$& $8.4$&$9.7$ & $8.8$&$9.2$\\
\hline
\multicolumn{8}{|c|}{\sc Resulting $\Trh$ and $\Yg$}\\\hline
$\Trh/10^{9}~\GeV$&$1.2$&$0.89$& $0.89$& $0.99$&$0.91$ & $0.99$&$1.2$\\
$10^{13}\Yg$&$2.4$&$1.7$& $1.7$& $1.9$&$1.7$ & $1.88$&$1.88$\\
\hline
\end{tabular}
\caption{\small\sl Parameters yielding the correct BAU for various
neutrino mass schemes for $\ld=0.01$ and $\ck=220$.}
\label{tab4}\ec
\end{table}

In this part of our numerical program, for a given neutrino mass
scheme, we take as input parameters: $\mn[\ell], \mD[i],
\varphi_1, \varphi_2$ and the best-fit values of the neutrino
parameters listed in \Tref{tabn}. We then find the
\emph{renormalization group} (RG) evolved values of these
parameters at the scale of nTL, $\Lambda_L$, which is taken to be
$\Lambda_L=\msn$, integrating numerically the complete expressions
of the RG equations -- given in \cref{running} -- for $\mn[i]$,
$\theta_{ij}$, $\delta$, $\varphi_1$ and $\varphi_2$. In doing
this, we consider the MSSM with $\tan\beta\simeq50$, favored by
the preliminary LHC results -- see, e.g., \cref{shafi,lpnova1} --
as an effective theory between $\Lambda_L$ and a SUSY-breaking
scale, $M_{\rm SUSY}=1.5~\TeV$. Following the procedure described
in \Sref{lept2}, we evaluate $\mrh[i]$ at $\Lambda_L$. We do not
consider the running of $\mD[i]$ and $\mrh[i]$  and therefore, we
give their values at $\Lambda_L$.

We start the exposition of our results arranging in \Tref{tab4}
some representative values of the parameters leading to the
correct BAU for $\ld=0.01$ and $\ck=220$ and normally hierarchical
(cases A and B), degenerate (cases C, D and E) and invertedly
hierarchical (cases F and G) $\mn[]$'s. For comparison we display
the $B$-yield with ($Y_B$) or without ($Y^0_B$) taking into
account the RG effects. We observe that the two results are more
or less close with each other. In all cases the current limit of
\Eref{sumn} is safely met -- the case D approaches it --, while
$m_\beta$ turns out to be well below the projected sensitivity of
KATRIN \cite{katrin}. Shown are also the obtained $\Trh$'s, which
are close to $10^9~\GeV$ in all cases, and the corresponding
$Y_{\Gr}$'s, which are consistent with \Eref{Ygw} for
$m_{\Gr}\gtrsim11~\TeV$.

\begin{figure}[!t]\vspace*{-.15in}
\hspace*{-.25in}
\centering
\includegraphics[height=2.27in]{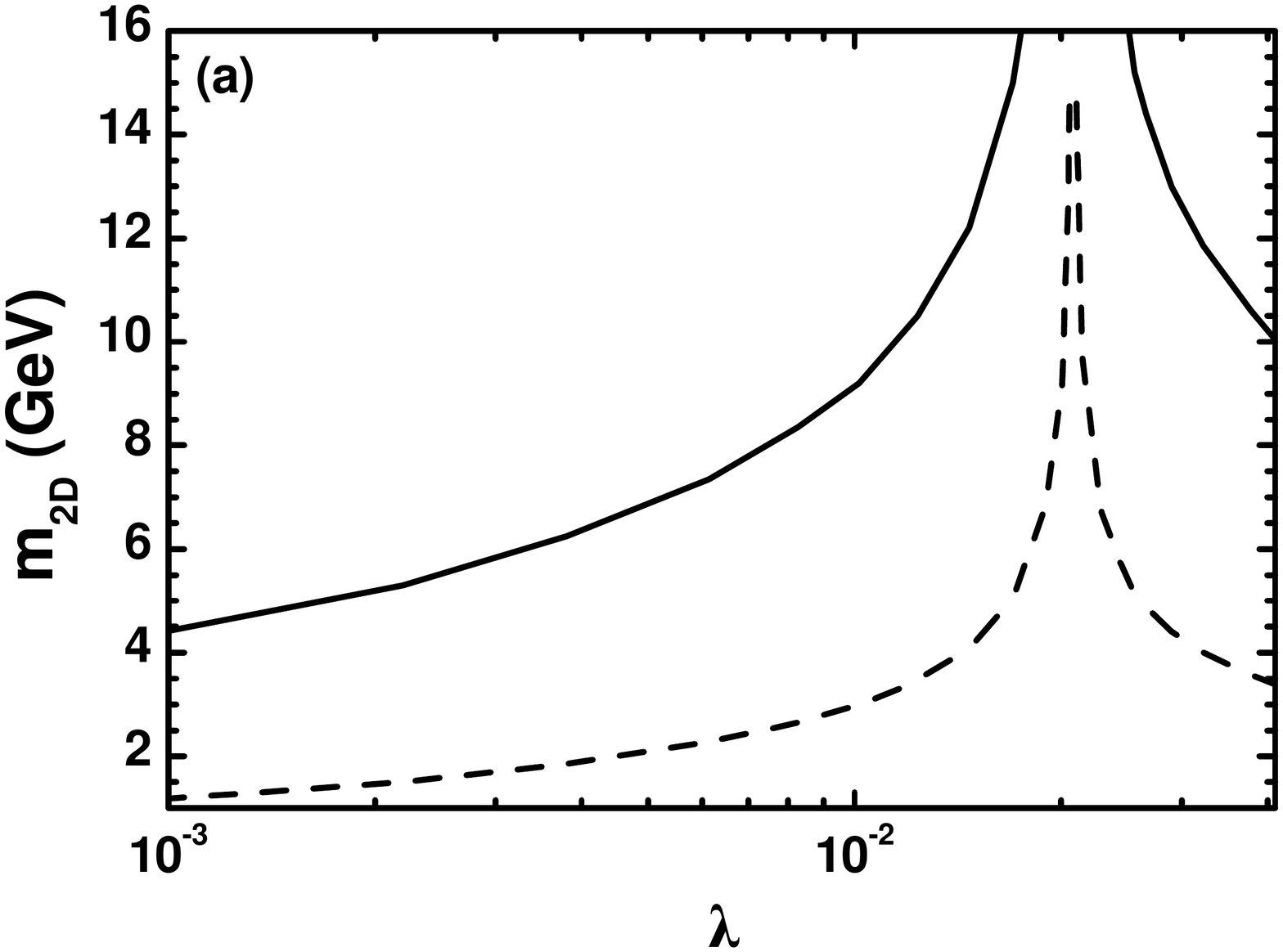}\hspace*{-0.8cm}
\includegraphics[height=2.27in]{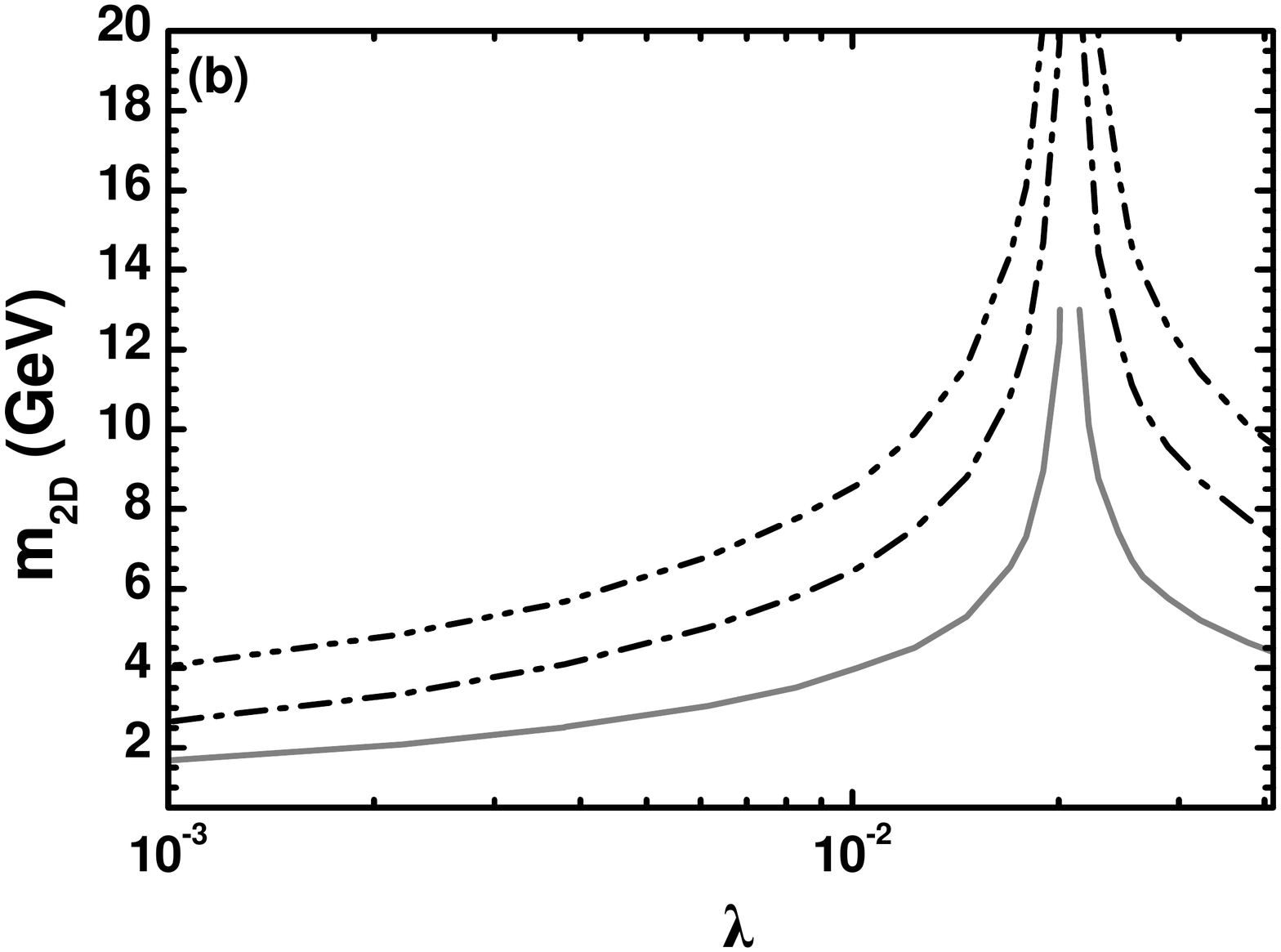}\hfill
\\\vspace*{-.4in}\hspace*{-.25in} \centering
\includegraphics[height=2.27in]{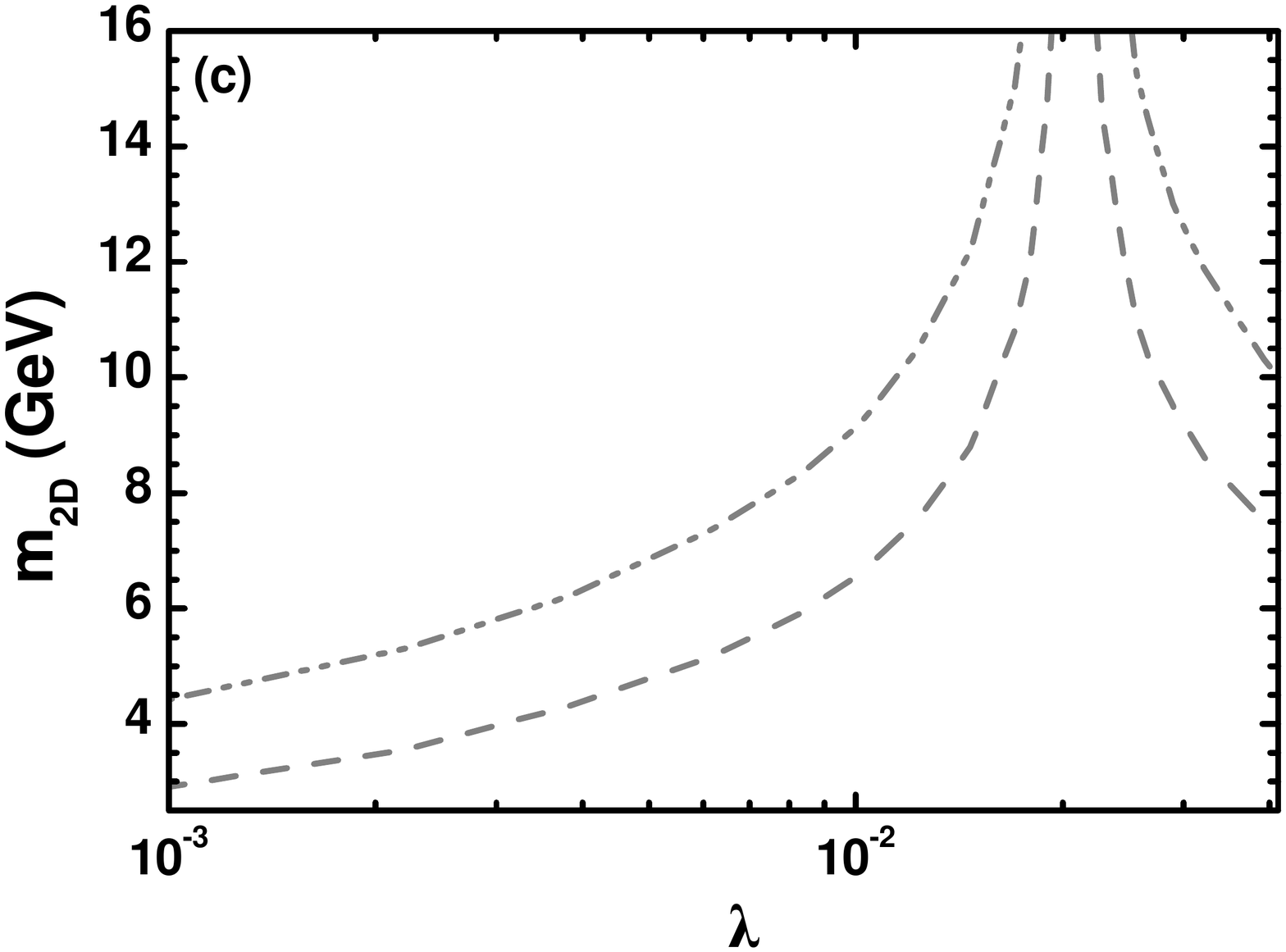}\hspace*{-0.8cm}
\includegraphics[height=2.27in]{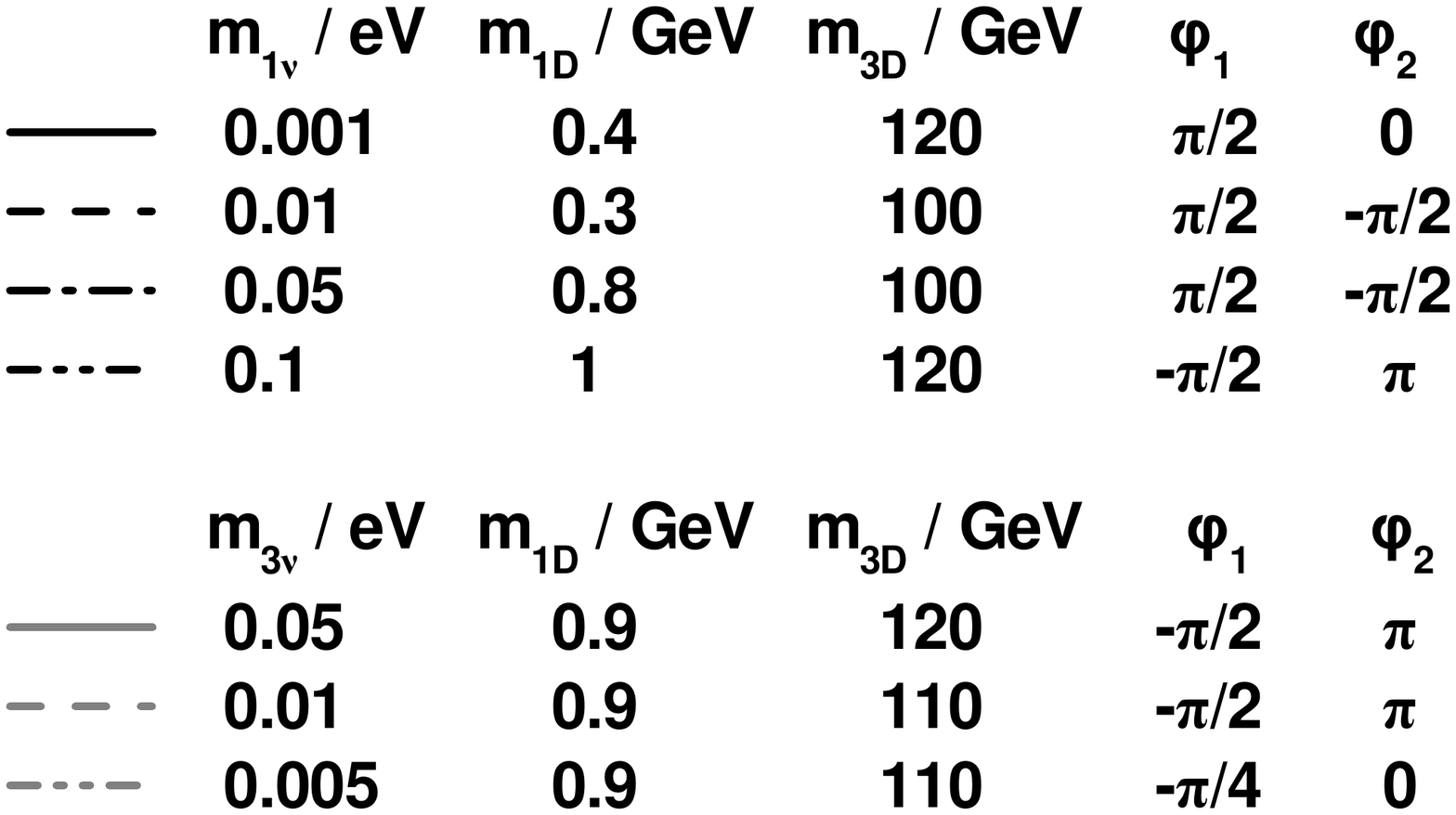}\hfill\\\vspace*{-.3in}
\renewcommand{\arraystretch}{1.}
\begin{center} \begin{tabular}{|c||c|c||c|c|c|c|} \hline
%
$\mn[1] (\eV)$&$\ld~(10^{-2})$&$\mD[2] (\GeV)$&$\mrh[1] (10^{10}~\GeV)$&$\mrh[2] (10^{12}~\GeV)$&$\mrh[3] (10^{14}~\GeV)$\\
\hline
%
%
$0.001$&$0.1-1.8$&$5.5-17$&$4.7$&$1-9.8$&$32$\\&$2.4-4.1$&$17-10$&$4.7$&$17-10$&$32$\\
$0.01$&$0.1-2$&$1.5-13$&$3.5-6$&$0.26-8.75$&$0.33-0.4$\\
&$2-4.1$&$15-3.4$&$6-5.3$&$11-0.84$&$0.42-0.34$\\
$0.05$&$0.1-2.$&$3-28$&$3.5-4$&$2.1-11$&$0.5-0.6$\\
&$2.1-4.1$&$28-7.3$&$4-3.9$&$11-0.8$&$0.6-0.5$\\
$0.1$&$0.1-1.9$&$4-23$&$1.3$&$0.4-8$&$0.4-0.5$\\
&$2.2-4.1$&$23-9$&$1.3$&$10-2$&$0.5-0.4$\\\hline
%
$\mn[3] (\eV)$&$\ld~(10^{-2})$&$\mD[2] (\GeV)$&$\mrh[1]
(10^{10}~\GeV)$&$\mrh[2] (10^{12}~\GeV)$&$\mrh[3]
(10^{14}~\GeV)$\\\hline
$0.05$&$0.1-2$&$2.1-12$&$1.5-1.6$&$0.28-7.8$&$0.48-0.56$\\
&$2.1-4.1$&$13-4.4$&$1.6$&$8.7-1.2$&$0.57-0.49$\\
$0.01$&$0.1-2$&$2.8-17$&$2.2$&$0.4-11$&$5.4-5.6$\\
&$2-4.1$&$17-7.3$&$2.2$&$11-2$&$5.6-5.4$\\
$0.005$&$0.1-1.8$&$5-17$&$1.8$&$0.7-10.1$&$12$\\
&$1.8-4.1$&$17.7-10$&$1.8$&$10.5-3.5$&$12$\\\hline
\end{tabular}\end{center}\hfill \vspace*{-.1in}
\caption{\sl\small Contours on the $\ld-\mD[2]$ plane, yielding
the central $Y_B$ in \Eref{BAUwmap}, consistently with the
inflationary requirements, for $\lH=\lHb=0.5, \kx=1$ and
$y_{33}=0.6$ and various ($\mn[\ell], \mD[1], \varphi_1,
\varphi_2)$'s indicated next to the graph (c) and NO [IO]
$\mn[i]$'s (black [gray] lines). The corresponding ranges of
$\mrh[i]$'s are also shown in the table included. }\label{fig2}
\end{figure}


From \Tref{tab4} we also remark that the achievement of $Y_B$
within the range of \Eref{BAUwmap} dictates a clear hierarchy
between the $\mrh[i]$'s, which follows the imposed hierarchy in
the sector of $\mD[i]$'s -- see paragraph 6 of \Sref{cont1}. This
is expected since, in the limit of hierarchical $\mD[i]$'s, the
$\mrh[i]$'s can be approximated by the following expressions
\cite{frigerio,branco}
\beqs\beq  \label{Mra}
\lf\mrh[1],\mrh[2],\mrh[3]\rg\sim\left\{\bem
\lf{\mD[1]^2\over m_{2\nu}s^2_{12}}\,,{2\mD[2]^2\over
m_{3\nu}}\,,{\mD[3]^2s^2_{12}\over 2m_{1\nu}}\rg &~\mbox{for NO
$\mn[]$'s} \cr
\lf{\mD[1]^2\over\sqrt{|\Delta m_{31}}|}\,,
{2\mD[2]^2\over\sqrt{|\Delta m_{31}}|}\,,{\mD[3]^2\over
2m_{3\nu}}\rg & \mbox{for IO $\mn[]$'s} \cr\eem
\right.\eeq Indeed, we see e.g. that for fixed $j$, the
$\mrh[j]$'s depends exclusively on the $\mD[j]$'s and $\mrh[3]$
increases when $\mn[\ell]$ decreases with fixed $\mD[3]$. As a
consequence, satisfying \sEref{kin}{a} pushes the $\mD[1]$'s well
above the mass of the quark of the first generation. Similarly,
the $\mD[2]$'s required by \Eref{BAUwmap} turns out to be heavier
than the quark of the second generation. Also, the required by
seesaw $\mrh[3]$'s are lower in the case of degenerate $\nu_i$
spectra and can be as low as $3\cdot 10^{13}~\GeV$ in sharp
contrast to our findings in \cref{nmH}, where much larger
$\mrh[3]$'s are necessitated. An order of magnitude estimation for
the derived $\eL$'s can be achieved by \cite{frigerio,branco} \beq
\label{eLa} \ve_2\sim -{3\mrh[2]\over 8\pi\vev{H_u}^2}\left\{\bem
\lf\mn[1]\over s_{12}^2\rg \hfill   & \mbox{~for NO $\mn[]$'s} \cr
\mn[3]\hfill   & \mbox{for IO $\mn[]$'s} \cr\eem
\right.\eeq\eeqs which is rather accurate, especially in the case
of IO $\mn[]$'s.

To highlight further our conclusions inferred from \Tref{tab4}, we
can fix $\mn[\ell]$ ($\mn[1]$ for NO $\mn[i]$'s or $\mn[3]$ for IO
$\mn[i]$'s) $\mD[1]$, $\mD[3]$, $\varphi_1$ and $\varphi_2$ to
their values shown in this table and vary $\mD[2]$ so that the
central value of \Eref{BAUwmap} is achieved. This is doable since,
according \Eref{Mra}, variation of $\mD[2]$ induces an exclusive
variation to $\mrh[2]$ which, in turn, heavily influences $\ve_L$
-- see \eqs{el}{hieF} -- and $Y_L$ -- see \eqs{Yb}{Gmqq}. The
resulting contours in the $\ld-\mD[2]$ plane are presented in
\Fref{fig2} -- since the range of \Eref{BAUwmap} is very narrow
the possible variation of the drawn lines is negligible. The
resulting $\mrh[j]$'s are displayed in the table included. The
conventions adopted for the types and the color of the various
lines are also described next to the graph (c) of \Fref{fig2}. In
particular, we use black [gray] lines for NO [IO] $\mn[i]$'s. The
black dashed and the solid gray line terminate at the values of
$\mD[2]$ beyond which \Eref{BAUwmap} is non fulfilled due to the
violation of \sEref{kin}{b}.

In all cases, two disconnected allowed domains arise according to
which of the two contributions in \Eref{l1} dominates. The
critical point $(\ld_{\rm c}, c_{\cal R\rm c})$ is extracted from:
\beq 1-12c_{\cal R\rm c}\mpq^2=0\Rightarrow c_{\cal R\rm c}=c^{\rm
max}_{\cal R}/2\simeq453\>\>\mbox{or}\>\>\ld_{\rm
c}\simeq10^{-4}c^{\rm max}_{\cal R}/4.2\simeq0.021\eeq
where we make use of \Eref{Mpqf} and \Eref{lan} in the
intermediate and the last step respectively. From
\eqss{T1rh}{Yb}{Gmqq} one can deduce that for $\ld<\ld_c$, $\Trh$
remains almost constant; $\GNsn[2]/\Gsn$ decreases as $\ck$
increases and so the $\mrh[2]$'s, which satisfy \Eref{BAUwmap},
increase. On the contrary, for $\ld>\ld_c$, $\GNsn[2]/\Gsn$ is
independent of $\ck$ but $\Trh$ increases with $\ck$ and so the
fulfilling \Eref{BAUwmap} $\mrh[2]$'s decrease.

Summarizing, we conclude that our scenario prefers the following
ranges for the $\mrh[i]$'s: \beqs\beq
1\lesssim\mrh[1]/10^{10}~\GeV\lesssim6,\>\>0.6\lesssim\mrh[2]/10^{12}~\GeV\lesssim20
,\>\>0.3\lesssim\mrh[3]/10^{14}\GeV\lesssim30,\eeq while the
$\mD[1]$ and $\mD[2]$ are restricted in the ranges: \beq
0.3\lesssim\mD[1]/\GeV\lesssim1,\>
1.5\lesssim\mD[2]/\GeV\lesssim20.\eeq \eeqs

\section{Conclusions}\label{con}

We investigated the implementation of nTL within a realistic GUT,
based on the PS gauge group. Leptogenesis follows a stage of nMHI
driven by the radial component of the Higgs field, which leads to
the spontaneous breaking of the PS gauge group to the SM one with
the GUT breaking v.e.v identified with the SUSY GUT scale and
without overproduction of monopoles. The model possesses also a
resolution to the strong CP and the $\mu$ problems of the MSSM via
a PQ symmetry which is broken during nMHI and afterwards. As a
consequence the axion cannot be the dominant component of CDM, due
to the present bounds on the axion isocurvature fluctuation.
Moreover, we briefly discussed scenaria in which the potential
axino and saxion overproduction problems can be avoided.

Inflation is followed by a reheating phase, during which the
inflaton can decay into the lightest, $\wrhn[1]$, and the
next-to-lightest, $\wrhn[2]$, RH neutrinos allowing, thereby for
nTL to occur via the subsequent decay of $\wrhn[1]$ and
$\wrhn[2]$. Although other decay channels to the SM particles via
non-renormalizable interactions are also activated, we showed that
the production of the required by the observations BAU can be
reconciled with the observational constraints on the inflationary
observables and the $\Gr$ abundance, provided that the (unstable)
$\Gr$ masses are greater than $11~{\rm TeV}$. The required by the
observations BAU can become consistent with the present low energy
neutrino data, the restriction on $\mD[3]$ due to the PS gauge
group and the imposed mild hierarchy between $\mD[1]$ and
$\mD[2]$. To this end, $\mD[1]$ and $\mD[2]$ turn out to be
heavier than the ones of the corresponding quarks and lie in the
ranges $(0.1-1)~\GeV$ and $(2-20)~\GeV$ while the obtained
$\mrh[1]$, $\mrh[2]$ and $\mrh[3]$  are restricted to the values
$10^{10}~\GeV$, $\lf10^{11}-10^{12}\rg~\GeV$ and
$\lf10^{13}-10^{15}\rg~\GeV$ respectively.

\section*{Acknowledgements} We would like to thank A.B. Lahanas, G.
Lazarides and V.C. Spanos for valuable discussions.

\def\ijmp{{\sl Int. Jour. Mod. Phys.}}
\def\plb{{\sl Phys. Lett. B }}
\def\prl{{\sl Phys. Rev. Lett.}}
\def\rmp#1#2#3{{Rev. Mod. Phys.}
{\bf #1},~#3~(#2)}
\def\prep{{ Phys. Rep. }}
\def\prd{{Phys. Rev. D }}
\def\npb{{Nucl. Phys. {\bf B}}}
\def\npps#1#2#3{{Nucl. Phys. B (Proc. Sup.)}
{\bf #1},~#3~(#2)}
\def\mpl#1#2#3{{Mod. Phys. Lett.}
{\bf #1},~#3~(#2)}
\def\jetp#1#2#3{{JETP Lett. }{\bf #1}, #3 (#2)}
\def\app#1#2#3{{Acta Phys. Polon.}
{\bf #1},~#3~(#2)}
\def\ptp#1#2#3{{Prog. Theor. Phys.}
{\bf #1},~#3~(#2)}
\def\n#1#2#3{{Nature }{\bf #1},~#3~(#2)}
\def\apj#1#2#3{{Astrophys. J.}
{\bf #1},~#3~(#2)}
\def\mnras#1#2#3{{MNRAS }{\bf #1},~#3~(#2)}
\def\grg#1#2#3{{Gen. Rel. Grav.}
{\bf #1},~#3~(#2)}
\def\s#1#2#3{{Science }{\bf #1},~#3~(#2)}
\def\ibid#1#2#3{{\it ibid. }{\bf #1},~#3~(#2)}
\def\cpc#1#2#3{{Comput. Phys. Commun.}
{\bf #1},~#3~(#2)}
\def\astp#1#2#3{{Astropart. Phys.}
{\bf #1},~#3~(#2)}
\def\epjc{{Eur. Phys. J. C}}
\def\jhep{{J. High Energy Phys.}}
\newcommand\jcap{J.\ Cosmol.\ Astropart.\ Phys.}
\newcommand\njp{New.\ J.\ Phys.}
\newcommand{\hepth}[1]{{\tt hep-th/#1}}
\newcommand{\hepph}[1]{{\tt hep-ph/#1}}
\newcommand{\hepex}[1]{{\tt hep-ex/#1}}
\newcommand{\astroph}[1]{{\tt astro-ph/#1}}
\newcommand{\arxiv}[1]{{\tt  arXiv:#1}}


\end{document}